\documentclass[12pt,preprint]{aastex}
\begin{document}

\parindent=1.0cm

\title{The Young Outer Disk of M83 \altaffilmark{1}}

\author{T. J. Davidge}

\affil{Herzberg Institute of Astrophysics,
\\National Research Council of Canada, 5071 West Saanich Road,
\\Victoria, BC Canada V9E 2E7\\ {\it email: tim.davidge@nrc.ca}}

\altaffiltext{1}{Based on observations obtained at the
Gemini Observatory, which is operated by the Association of Universities
for Research in Astronomy, Inc., under a co-operative agreement with the
NSF on behalf of the Gemini partnership: the National Science Foundation
(United States), the Science and Technology Facilities Council
(United Kingdom), the National Research Council of Canada (Canada), CONICYT (Chile),
the Australian Research Council (Australia), the Ministerio da Ciencia e Technologia (Brazil),
and the Ministerio de Ciencia, Tecnologia e Innovacion Productiva (Argentina).}

\begin{abstract}

	Deep near-infrared images recorded with NICI on Gemini South are used to 
investigate the evolved stellar content in the outer regions of the south east quadrant of the 
spiral galaxy M83. A diffuse population of asymptotic giant branch (AGB) stars is detected, 
indicating that there are stars outside of the previously identified 
young and intermediate age star clusters in the outer disk. The brightest AGB stars have M$_K 
\geq -8$, and the AGB luminosity function (LF) is well-matched by model LFs that assume 
ages $\leq 1$ Gyr. The specific SFR during the past few Gyr estimated from AGB star counts is 
consistent with that computed from mid-infrared observations of star clusters at similar 
radii, and it is concluded that the disruption timescale for star clusters in the outer disk 
is $<< 1$ Gyr. The luminosity function and specific frequency of AGB stars varies with radius, in 
a manner that is indicative of lower luminosity-weighted ages at larger radii. 
Modest numbers of red supergiants (RSGs) are also found, indicating that there has been star 
formation during the past 100 Myr, while the ratio of C stars to M giants 
is consistent with that expected for a solar metallicity system that has experienced 
a constant star formation rate (SFR) for the past few Gyrs. 
The results drawn from the properties of resolved AGB stars are 
broadly consistent with those deduced from integrated light observations in the UV. 

\end{abstract}

\keywords{galaxies: individual {M83} -- galaxies: evolution -- galaxies: spiral}

\section{INTRODUCTION}

	The angular momentum distribution of material within in a protogalactic halo affects the 
mechanics and timescale of structure formation in that halo (e.g. van den Bosch et al. 2002). 
The steady accretion of gas with progressively higher angular momentum with time causes 
disks to form from the inside--out, and the rate of growth will depend on the angular 
momentum distribution in the host halo. Simulations in which gas in an 
isolated, spherical dark matter-dominated halo undergoes a dissipative collapse produce a disk 
that grows throughout the age of the Universe at a rate of 
$\approx 0.7$ kpc Gyr$^{-1}$ (e.g. Roskar et al. 2008). 

	A prolonged period of disk assembly is consistent with observations. The 
most direct evidence of outward disk growth comes from studies of the size evolution of 
disks over a range of redshifts (e.g. Trujillo \& Agerri 2004; Trujillo \& Pohlen 2005). 
Age and metallicity gradients are also a natural consequence of inside--out disk formation, 
with the oldest luminosity-weighted ages and most chemically enriched material expected at 
small radii. Bell \& De Jong (2000) find evidence for such systematic gradients in nearby 
galaxies, while the outer regions of disks in a significant fraction of nearby spirals 
have an ultraviolet (UV) spectral-energy distribution (SED) that signals relatively 
young (at least in relation to smaller radii) luminosity-weighted ages (e.g. Zaritsky 
\& Christlein 2007). Looking at galaxies within 
the Local Group, significant amounts of infalling gas are 
required to model the chemo-photometric properties of the outer regions of M33. The rate of 
gas accretion in that galaxy appears to have peaked 3 -- 7 Gyr in the past, with more subdued 
accretion rates during the past 3 Gyr (Barker \& Sarajedini 2008). 

	Changes in disk properties are not spurred exclusively by the accretion of new material. 
Secular processes re-distribute angular momentum between stars and gas, and 
objects that gain angular momentum migrate to larger radii (e.g. Sellwood \& Binney 
2002), pushing the disk boundary even further outwards (Aguerri, Balcells, \& Peletier 2001). 
Galaxies also do not evolve in isolation, and interactions with companions will mix the 
stellar content of disks, changing the amplititude of (or even completely obliterating) 
radial trends in disk properties that were imprinted early-on, while also 
altering the thickness of the disk (e.g. Hayashi \& Chiba 2006). Major interactions 
also extend the timescale of disk assembly, as the net angular momentum content of surviving 
halos grows with time (e.g. D'Onghia et al. 2006). Evidence that the evolution of a typical disk 
galaxy is affected by mergers -- even during recent epochs -- comes from the morphologies 
of galaxies at intermediate redshift (Hammer et al. 2005) and the angular momentum content of 
nearby disk galaxies (Hammer et al. 2007), which together suggest that the majority of nearby 
spiral galaxies experienced significant mergers in the last few Gyr. Still, that the 
star-forming histories of massive spiral galaxies at intermediate epochs can be represented 
by models that assume passive evolution suggests that the impact of mergers is dampened 
on timescales that are substantially shorter than the Hubble time (Buat et al. 2008).

	Studies of resolved stars in the outer regions of nearby spiral galaxies 
are of key importance for furthering our understanding of 
disk evolution, and the Centaurus Group (CG) contains prime 
targets for detailed investigation. Many galaxies in the CG show signs of on-going or recently 
elevated SFRs, signalling that this group may be in a state of active dynamical evolution. 
One such galaxy is M83, which is the largest spiral in the CG, and is the center of one of 
the two galaxy concentrations in the CG (Karachentsev et al. 2002). M83 is 
experiencing starburst activity (Bohlin et al. 1983), with 
a present-day SFR of $3 - 4$ M$_{\odot}$ year$^{-1}$ (Boissier et al. 2005). 

	The bulk of the HI in M83 is in a thin disk, although there 
is a kinematially distinct, thick HI disk containing 
5.5\% of the total HI within the central 8 kpc (Miller, Bergman, \& Wakker 2009). 
Still, the bulk of the interstellar medium (ISM) in the inner disk of M83 is in molecular 
form, accounting for 80\% of the total ISM mass in the central 5 arcmin (Crosthwaite et 
al. 2002). The projected densities of CO and HI drop markedly at a radius of 5.5 arcmin 
(Crosthwaite et al. 2002). The drop in molecular gas content at large radii may partially 
be a consequence of low ISM pressure, which hinders the formation of molecular material 
(Crosthwaite et al. 2002). The decline in gas density is 
accompanied by warping of the disk (Rogstad, Lockhart, \& Wright 1974; Huchtmeier \& 
Bohnenstegel 1981), and the disappearance of prominent spiral structure. 

	The drop in the interstellar content at 5.5 arcmin does not herald the 
termination of the gas disk; rather, the gas disk of M83 
extends well past the classical optical radius (Rogstad et al. 1974), and Huchtmeier 
\& Bohnenstengel (1981) conclude that 80\% of the total HI mass lies outside of the Holmberg 
radius. The HI at large radius may not be in a thin disk, as 
kinematic measurements suggest that the vertical scale height of stars in the 
outer disk of M83 flares, in a manner than is consistent with heating by interactions with 
sub-structures in the halo (Herrmann, Ciardullo, \& Sigurdsson 2009). Still, 
the ISM reservoir at large radius in M83 has been the site of recent star formation, 
as observations made with GALEX detect UV light from diffuse and compact components at 
large radii (Thilker et al. 2005). The compact UV sources are star clusters and HII regions 
that coincide with peaks in the HI distribution where the ISM is near the local 
critical density for triggering star formation (Dong et al. 2008). The clusters in the 
outer disk have typical masses log(M$_{\odot}$) = 4.7, with a substantial 
dispersion (Dong et al. 2008). While the mass distribution of clusters in M83 is similar to 
that in the outer disks of other spiral galaxies, the UV light in M83 
extends to larger radii than in these other systems (Zaritsky \& Christlein 2007).

	Star formation in the outer regions of M83 has not been an exclusively recent 
phenomenon. The spectra of many outer disk clusters lack H$\alpha$ emission, suggesting ages 
in excess of 10 Myr. Ages estimated from spectral-energy distributions (SEDs) indicate 
that the majority of clusters have ages $\leq 350$ Myr, and Dong et al. 
(2008) conclude that cluster formation has been on-going for at least 1 Gyr after 
accounting for selection effects that hinder the detection of older systems. Finally, HII 
regions in the outer regions of M83 have metallicities that are at least one third 
solar, and this is consistent with the current levels of star formation 
having been sustained over Gyr or longer time scales if there has been 
{\it in situ} enrichment (Bresolin et al. 2009). This being said, feedback from SNe and 
other secular processes may also re-distribute chemically enriched material throughout 
the disk, boosting the metallicity of peripheral disk regions to higher levels than if they 
had evolved as a closed system (e.g. Scannapieco et al. 2008).

	The local SFR in M83 has varied with radius during recent epochs. The luminosity 
functions (LFs) of x-ray sources in the central regions and disk of M83 differ (Soria \& Wu 
2002), such that the nucleus appears to have experienced a more continuous recent star-forming 
history than the main body of the disk, where the star-forming activity has recently dropped.
This trend of decreasing age with increasing radius does not extend 
to very large radii, where the azimuthally-averaged UV color of M83 becomes bluer with 
increasing radius, and a rich population of compact young clusters is seen (Thilker et al. 2005). 
The clusters in the outer disk of M83 are smaller than those in the classical 
disk, and the brightest outer disk clusters are 2 - 3 mag fainter in the FUV than their 
inner disk kin (Bresolin et al. 2009). The line emission in outer disk HII regions is 
typically powered by a solitary O star, as opposed to the $\sim 10^2$ such stars 
that power HII regions at smaller radii (Bresolin et al. 2009). 

	The inner and outer regions of the M83 disk may also have experienced 
systematically different chemical evolution histories. While the analysis of spectra from 
HII regions indicate that there is an [O/H] gradient inside the Holmberg radius of M83, this 
trend does not continue to larger radii (Bresolin et al. 2009). Bresolin et al. (2009) argue 
that this chemical homogeneity may be a consequence of the low gas density at large radii. 
Drawing a parallel with low surface brightness (LSB) galaxies, which do not show large radial 
abundance gradients because star formation occurs sporatically, Bresolin et al. (2009) argue 
that a similar mechanism may be at work in the outer regions of M83. 
The material in the outer disk of M83 may also have been churned by tidal forces.

	In the present study, stars that are evolving on the asymptotic giant branch (AGB) are 
used to probe the star-forming history of the outer disk of M83 during the past 
few Gyr. These stars are identified in deep near-infrared images obtained with 
NICI on Gemini South (GS), and this is the first investigation in which 
individual stars are resolved in the outer regions of this galaxy. The adaptive optics (AO) 
capabilities of NICI, coupled with the ability to image at near-infrared wavelengths, make 
this instrument a powerful tool for probing the cool stellar contents of nearby galaxies. 
Because of their cool temperatures and the impact of line blanketing on the 
visible spectral-energy distributions (SEDs) of even moderately metal-rich evolved stars, 
AGB and RGB stars are arguably best studied at near-infrared wavelengths. 
Studies of resolved stars also benefit greatly from the high angular 
resolution delivered by NICI. Finally, background galaxies are the dominent source of 
contamination in deep photometric surveys of this nature, and the $\sim 0.1$ arcsec 
angular resolution of these images is helpful for identifying these objects. 

	The paper is structured as follows. The observations and photometric measurements are 
discussed in \S 2, while the color-magnitude diagrams (CMDs) and luminosity functions (LFs) 
constructed from these data are discussed in \S 3. The spatial distribution of AGB and 
RGB stars is investigated in \S 4. A discussion and summary of the results follows in \S 5. 

\section{OBSERVATIONS, DATA REDUCTION, AND PHOTOMETRIC MEASUREMENTS}

\subsection{The NICI Observations}

	Deep near-infrared images of five fields in the outer disk of M83 were obtained with 
NICI (Artigau et a. 2008) during March and April 2009 as part of queue program GS2009A-Q-74; 
these were the first queue-scheduled observations made with this instrument. NICI uses natural 
guide stars (NGSs) as beacons for AO correction. The NICI AO system assumes that the dominant 
turbulent layer is at the same elevation as the primary mirror. A beamsplitter divides the 
light into a `blue' and a `red' light path. While a suite of beamsplitters will ultimately 
be deployed, at the time of these observations only the `50--50 $H-$band' beamsplitter, 
which parses light into wavelengths longward and shortward of $1.85\mu$m, 
was installed. The detector in each light path is a $1024 \times 1024$ InSb 
array, with an angular scale of 0.018 arcsec pixel$^{-1}$; hence, each detector images 
$18 \times 18$ arcsec$^2$ on the sky. 

	Each field is centered on a star with $V < 14$, which serves as the NGS. 
The locations of the NGSs, which are indicated on the Digital Sky Survey image in Figure 1, 
allow a range of galactocentric radii in the outer south 
east quadrant of the galaxy to be explored. The right hand panel of Figure 1 shows 
the map of 1.6 cm emission from Tilanus \& Allen (1993), and three of the NGSs 
sample parts of the sky that are either in or near prominent HI concentrations. The 
field centered on GSC06726-0259 is near a region of pronounced FUV emission.
Additional information about the guide stars is listed in Table 1. R$_{GC}$ 
is the distance from the center of M83 measured in the disk plane, assuming (1) a 
distance modulus of 28.6 (Karachentseva et al. 2007), and (2) a position angle and disk 
inclination determined from the $K-$band image of M83 in the 2MASS Large Galaxy 
Atlas (Jarrett et al. 2003). 

	Most of the images used in this study were recorded 
simultaneously through $H$ and $Ks$ filters. With the exception of one 
field (see below), a single observation consisted of five 30 sec co-adds. Observations were 
obtained at five points along a linear dither pattern with 1 arcsec offsets. This dither 
pattern was employed to facilitate the suppression of cosmic-rays and the construction of the 
calibration frames that are required to remove interference 
fringes and the thermal signatures of warm objects along the optical path. The observing 
sequence was repeated multiple times to boost the S/N ratio of the final dataset, and the 
total exposure time is $\approx 2$ hours field$^{-1}$ filter$^{-1}$.

	The field centered on GSC 06726--00126 was the first to be observed, and 
these observations differed from those of the other fields, as the knowledge gained 
from these data was used to refine the overall observing strategy. 
The initial plan was to record images simultaneously in $J$ and $K_s$. However, the 
GSC 06726--00126 observations revealed that the throughput of the 
50--50 $H-$band beamsplitter in $J$ is markedly lower than anticipated, and so data were 
recorded in $H$ instead of $J$ for the remaining fields. A series of 
$J$ and $H$ observations were subsequently obtained of the GSC 06726--00126 field 
with NICI in single channel mode, in which a mirror feeds all of the light to only 
one channel. The $H-$band photometry of this field is not as deep as for the other fields.

	The initial processing of the images followed standard procedures. The first steps 
were dark-subtraction and the division by a flat field frame. The flat field frame was 
obtained from observations of a light source in the facility cassegrain 
calibration unit. A series of exposures were recorded with this source `on' and 
`off'. The latter data monitor thermal emission from warm objects along the light path, and a 
final flat field frame for each filter was constructed by subtracting the mean of the `off' 
exposures from the mean of the `on' exposures, and then normalizing the result.

	Interference fringes and signatures of warm objects along 
the optical path were removed by subtracting calibration 
frames that were constructed by median combining the science observations of each field 
in each filter. A DC sky level was subtracted from each observation prior to image 
combination to account for flucuations in the background light. 
The point spread function (PSF) wings from the NGSs extend over a few arcsec, and 
these produce conspicuous artifacts in the calibration frames if they are not removed. 
A $0.45 \times 0.45$ arcsec running median filter was applied to the flat-fielded images 
to suppress the PSF wings, and the results were subtracted from the 
flat-fielded images. A 0.45 arcsec filter size was selected as it was found not to affect the 
central portions of the PSF, while at the same time rejecting much of the light outside of the 
first diffraction ring. These `wing-suppressed' images were combined to construct the 
thermal signature/interference fringe calibration frame for each field$+$filter combination. 

	The processed images of each field were aligned and the 
results were combined by taking the median intensity at each pixel location. The final 
step in the processing sequence was to trim the median-combined images to the area 
that is common to all exposures of a given field, typically leaving a 
$\sim 14 \times 14$ arcsec$^2$ area for photometric analysis. An example of the final 
processed data is shown in Figure 2, where the H-band image of GSC 06726-00259 is displayed. 
There is a region with a $\approx 1 - 2$ arcsec radius surrounding the NGS 
where speckle noise prevents the detection of faint objects. 
Point sources, which have FWHM $\approx 0.08$ arcsec, 
are clearly seen at distances in excess of 2 arcsec from 
the guide star, and these are stars in the outer regions of M83. 

\subsection{Photometric Measurements}

	Stellar brightnesses were measured with the PSF-fitting routine ALLSTAR (Stetson \& 
Harris 1988). PSFs were constructed using the DAOPHOT (Stetson 1987) {\it PSF} task, 
using sources selected based on brightness, compactness, and the absence of 
obvious neighbors. Faint objects close to PSF stars were removed in an iterative manner. 
The PSF extraction radius was 0.3 arcsec; while the PSF wings extend to 
larger radii, sky noise confounds efforts to trace this part of the PSF in all but the 
brightest stars. The radius for PSF-fitting was restricted to the central diffraction spike, 
where the contrast between the target and background sky signals is highest.

	The photometric measurements were culled to remove objects with (1) fitting 
uncertainties computed by ALLSTAR, $\epsilon$, that exceed 0.3 magnitudes, which excises objects 
for which photometry is problematic, (2) higher than average $epsilon$ values 
when compared with other objects of the same magnitude, and (3) that are non-circular, 
as gauged by the DAOPHOT {\it ROUND} parameter. The first step sets the faint limit of the data, 
while the second and third steps remove objects that have a non-stellar appearance 
and/or are in very crowded environments. Speckle noise prevents the detection of all 
but the brightest objects within $\approx 2$ arcsec of the NGS, and so stars 
in that region were not photometered. Objects that coincide with 
diffraction spikes were also deleted from the target list. 

	The light path towards any source in the field (except 
the guide star) does not sample the same atmospheric cross-section 
as the guide star, and this is referred to as `anisoplanicity'. Anisoplanicity can complicate 
the construction of the PSF from single NGS AO-corrected data, depending on factors such as 
the field of view, the order of correction, and the atmospheric turbulence profile. The impact 
of anisoplanicity increases with angular distance from the NGS, with the 
result that the PSF varies with distance from the NGS. The 
impact of anisoplanicity is modest on photometric measurements made 
over angular scales of many arcsec from images recorded at good 
ground-based observing sites (e.g. Davidge \& Courteau 1999). 

	A modest amount of anisoplanicity is evident in some of the images, and 
this is demonstrated in Figure 3, where composite $H-$band light profiles of stars 
in two radial distance intervals from GSC06726--00259 are compared. The PSF of sources 
with $r_{NGS} > 3.6$ arcsec, where $r_{NGS}$ is the distance from the NGS, is $\sim 8\%$ 
broader than that of sources with $r_{NGS} < 3.6$ arcsec; the radial differences in FWHM are 
smaller in the $K-$band data. The uncertainty in the FWHM measurements, estimated from 
bootstrap re-sampling, is $\pm 0.003$ arcsec, indicating that 
the FWHMs of the PSFs differ at roughly the $2\sigma$ level. 

	Two sets of photometric measurements were made for each field: one set used 
a fixed PSF for each filter, and the other a spatially variable PSF. In the latter case 
it was assumed that PSF variations across the field follow a 
quadratic polynomial. The fixed and variable PSFs were constructed from the 
same stars to prevent introducing possible systematic effects when comparing 
the results. The photometry obtained with the variable PSF fit was adopted only if it produced 
a CMD with a smaller dispersion than the results obtained with the fixed PSF.

	The CMDs of the GSC06726--0259 field generated from fixed and 
variable PSFs, which are compared in Figure 4, showed the greatest 
differences. The application of a variable PSF reduces the 
dispersion in the CMD, and similar results were found for the GSC06726--0266 field, but not 
for the remaining fields. Therefore, a variable PSF was used to construct the final CMDs 
of the GSC06726--249 and 266 fields, while a fixed PSF was used for the other fields.

	The photometric calibration was defined using the brightnesses and colors of the 
NGSs listed in the 2MASS Point Source Catalogue (Cutri et al. 2003). The NGSs are saturated 
in the long-exposure science observations, and so the calibration measurements were made from 
short exposure observations that were interleaved with the longer-exposure observations 
throughout the observing sequence. This strategy was adopted so that the short exposure 
calibration images sample roughly the same range of Strehl ratios as the science data.
The short exposure images were recorded with the AO system running.

	If the Strehl ratio exceeds $\sim 0.05 - 0.1$, then the center of the PSF is 
dominated by a prominent spike, the width of which roughly matches the telescope 
diffraction-limit (the agreement with the theoretical diffraction limit is not exact, as optical 
abberations and residuals in image motion that are not corrected due 
to the limited bandwidth of the AO system broaden this feature). This central spike sits on 
a much more extended component, that is modulated at intermediate radii by the 
Airy pattern, and is spread over an angular scale that is comparable to 
that of the `seeing' disk. Calibration measurements that rely only on information in the central 
diffraction spike are uncertain due to Strehl ratio variations, such that $\epsilon_{cal} 
\approx \Delta S$, where $\epsilon_{cal}$ is the uncertainty in the calibration in magnitudes, 
and $S$ is the Strehl ratio. While more robust calibration measurements can be obtained using 
very large apertures, the results may then be affected by crowding, or uncertainties in 
the flat-field. There are also pragmatic difficulties in translating measurements made 
with a large aperture to measurements made with a much smaller PSF-fitting radius. 

	For the current data, the calibration is based on photometry within the PSF radius 
(0.3 arcsec -- see above). The fraction of light outside of the 
0.3 arcsec radius will change with seeing, and this will introduce 
scatter in the frame-to-frame measurements, as well as in the width of the sequences on the CMDs.
This calibration procedure has been found to produce excellent agreement with measurements 
made from non-AO corrected data (e.g. Figure 4 of Davidge 2007). After accounting for 
beamsplitter throughput, the photometric zeropoints computed for NICI in this manner agree with 
those measured for Hokupa'a $+$ QUIRC at Gemini North to within a few tenths of a magnitude. 

	The final science images were constructed from data that were 
typically recorded over at least 2 hours, and so Strehl ratio variations 
that occur over times scales of many minutes or less tend to average out. 
In fact, the scatter in the $H$ and $K$ zeropoints during any given night was typically $< \pm 
0.1$ mag, and this sets an upper limit to the reliability of the calibration. 
Evidence that the calibration is reliable to within a few percent 
comes from the mean $H-K$ colors computed from the various fields, which agree 
to within $0.006 \pm 0.021$ magnitude (\S 3.1.2). 

\section{RESULTS}

\subsection{The $(K, H-K)$ CMDs}

\subsubsection{The General Appearance of the CMDs and the Impact of Anisoplanicity}

	The $(K, H-K)$ CMDs of the five fields
are shown in Figure 5. There are field-to-field differences in source density, 
as might be expected given the range in R$_{GC}$ (Table 1). 
The lowest stellar density is in the GSC06726--0067 field, which has the 
highest R$_{GC}$, while the GSC06726--0259 field, which is close to an area of prominent 
UV emission, may be deficient in objects at the bright end. 

	The number of objects in each field is modest, 
making it difficult to identify coherent trends. However, this changes when 
data from multiple fields are combined. A composite CMD of all fields is shown in the 
lower right hand panel of Figure 5, and objects with $K < 22$ define a locus that is 
populated predominantly by AGB stars (\S 3.1.2). The composite CMD also broadens and becomes 
more richly populated when $K > 22$, possibly due to the onset of the RGB. 

	GSC06726--00126 and GSC06726--0181 have the smallest R$_{GC}$ and are located on the 
sky close to the south east spiral arm of M83. These fields are thus most likely to 
have the highest levels of contamination from stars in the classical disk of M83, and
their stellar contents might then be expected to differ from those of the other fields that 
have larger R$_{GC}$. In fact, the bright sequences in the CMDs of these two fields appear to be 
tilted from vertical, whereas the bright sequences in the other three 
fields are close to vertical. This being said, the normal points of the fields with 
the smallest and largest R$_{GC}$ are not significantly different (\S 3.1.2).

	Anisoplanicity introduces scatter in the photometry; even though the photometry was 
obtained with a variable PSF in some fields, residual variations in the ability to track the 
PSF are undoubtedly present, and these are a source of scatterin the CMDs. One way to check for 
systematic effects due to PSF variations is to examine the CMDs of stars at different offsets 
from the guide star. The CMDs of objects in two radial intervals near GSC06726--0259 
are shown in the right hand panel of Figure 4. There is excellent agreement 
between the CMDs defined by the stars in the two radial intervals.
An examination of the CMDs of sources at various radial intervals in the other fields 
also did not find evidence for radial variations in either the mean color 
or breadth of the stellar sequence. These comparisons suggest that any systematic errors 
due to anisoplanicity are $\leq 0.1$ magnitude in all of the fields. 

	Random noise, coupled with spatial variations in the PSF that are not tracked by 
DAOPHOT, introduce scatter in the photometric measurements. Artificial star tests, in which 
scaled versions of the PSFs are added to the data, were run to estimate 
the dispersion arising from random noise. The artificial stars were assigned the color of the 
dominant stellar locus in each field. As with the photometry of real stars, artificial stars 
were only considered to be recovered if they were detected in both passbands. 

	An upper limit to the scatter that PSF variations contribute 
to the CMDs can be obtained by comparing the observed scatter 
with that predicted from the artificial star experiments. 
Only an upper limit to the scatter is obtained, as 
star-to-star variations in age and/or metallicity, coupled with stellar variability, 
broaden the stellar locus on CMDs. At $K = 21$, the scatter in $H-K$ predicted 
from the artificial star experiments agrees with that seen among M giants in fields 
GSC06726--0181 and 0259, indicating that PSF variations in these fields are not 
significant when compared with random noise at this magnitude. 
As for the GSC06726--0126 and 0266 fields, there is a residual dispersion 
of $\pm 0.1$ magnitude in $H-K$ at $K = 21$, indicating that PSF variations 
introduce scatter of no more than this amount. The situation changes 
at fainter magnitudes. At $K = 22$ the measured dispersion in 
$H-K$ exceeds that predicted from the artificial star experiments in all four fields. 
Given that the scatter introduced by PSF variations will not change with 
magnitude, then the increased scatter at $K=22$ indicates that there is a sustantial mix in 
the age and/or metallicity of stars at this magnitude in all fields.

\subsubsection{Comparisons with isochrones and the K/M giant AGB content}

	The $(M_K, H-K)$ CMDs of two field groupings, one consisting of GSC06726--0126 
and 0181 (`inner disk'), which are close together on the sky and have the smallest R$_{GC}$, 
and the other of the remaining fields (`outer disk'), are compared in Figure 6. 
A distance modulus of 28.6, as computed by Karachentsev et al. (2007) from the 
brightness of the RGB-tip, has been assumed. As for extinction, the Schlegel, Finkbeiner, 
\& Davis (1998) maps give A$_B = 0.28$ for M83. This is a lower limit to the 
total extinction, as it does not include dust that is internal to M83. In fact, the outer regions 
of at least some spiral galaxies appear to contain significant amounts of dust. Popescu et al. 
(2005) find that the ratio of FIR to UV flux in M101 decreases with increasing radius, and 
interpret this as a radial gradient in attentuation. In the case 
of M83, the $8\mu$m emission from the star-forming knots 
in the outer disk of M83 is predominantly stellar in origin, suggesting that there 
is little interstellar dust in these clusters (Dong et al. 2008). Still, there is 
evidence for significant dust extinction outside of these clusters. 
Boissier et al. (2005) compute radial extinction profiles for M83 using three 
different indicators: (1) the ratio of Balmer line strengths, (2) the ratio 
of the total IR flux to the FUV flux, and (3) the slope of 
FUV SED. All three estimators predict that the attenuation reaches a minimum of A$_{FUV} 
= 1.5$ magnitudes at a radius of 200 arcsec, before climbing with increasing radius out to 
at least 6 arcmin. These three extinction indicators agree
out to a radius of 250 arcsec, but at larger radii the Balmer line ratio predicts a much 
larger increase in extinction towards larger radii than the other two; 
in contrast, the extinction estimated from the slope of the UV 
SED may decrease at very large radii, although there are substantial error bars in these 
measurements. As a compromise between these two extremes, the extinction computed from 
the total-IR/FUV ratio with A$_{FUV} = 1.6$ magnitudes (the mean at large radii in 
Figure 2b of Boissier et al. 2005) is adopted here, and has been applied to the data points 
in Figure 6. This corresponds to A$_V = 1.4$ magnitudes and A$_K =$ 0.16 magnitudes. 

	To facilitate the comparison of any general trends, mean $H-K$ colors were computed 
in $\pm 0.25$ magnitude bins in $K$ in the inner and outer disk CMDs, and the results are 
shown in Figure 6. A $2.5\sigma$ rejection scheme was applied to suppress 
outliers, and stars with $(H-K)_0 > 0.45$, which are either C stars or background 
galaxies, were not included when computing the mean colors. 
The error bars show the standard error in the mean color. The mean difference between the normals 
computed for the inner and outer disk fields is $\Delta(H-K) = 0.006 \pm 0.021$ magnitudes, 
in the sense of outer -- inner disk color. Therefore, the mean colors of the 
inner and outer disk sequences are not significantly different.

	Isochrones with Z = 0.004 and Z = 0.019 from Girardi et al. (2002) are 
plotted in Figure 6. These isochrones span the range of metallicities inferred from 
[O/H] measurements of HII regions in the outer disk (e.g. Bresolin et al. 2009). Figure 4 of 
Bresolin et al. (2009) summarizes [O/H] measurements in the outskirts of M83, and there is 
considerable scatter due to calibration uncertainties. The majority of calibrations 
indicate that [O/H] in the outer disk of M83 is within $\sim 0.1 - 0.2$ dex of solar 
(Z = 0.010 -- 0.013 if oxygen is assumed to trace all metals). There is an [O/H] gradient 
in the disk of M83, and these same calibrations predict a super-solar metallicity for 
the inner disk of M83. There is one calibration that yields [O/H] values that are markedly 
lower than the others, and this predicts that the outer disk has
[O/H] that is 0.4 dex lower than solar (Z = 0.006 if oxygen traces all metals). 

	While detailed abundance studies of the chemical compositions of bright stars 
will ultimately provide the information needed to confirm the metallicity scale in M83, there is 
a hint that the youngest stars in the classical disk of M83 may have a metallicity 
that is higher than solar. The ratio of Wolf-Rayet stars with 
carbon dominant (WC) to nitrogen dominant (WN) atmospheres 
is higher in M83 than in the Galaxy, and this may be due to a higher metallicity 
in M83, and the consequent impact on stellar mass loss rates (Crowther et al. 2004; 
Hadfield et al. 2005). 

	Photometric variability in highly evolved AGB stars introduces uncertainties 
when comparing observations with models, and there are also uncertainties in the model 
physics. This being said, the isochrones indicate that many of the stars with M$_K < -6.8$ are 
evolving on the AGB, and that those with M$_K < -7.8$ have ages $\leq 10$ Gyr. 
The mean color of stars at M$_K = -8$ thus provides some insight into metallicity, since these 
stars should not have colors that place them above very old isochrones. The mean colors 
of stars in the three outer disk fields at M$_K = -8$ coincides with the 10 Gyr Z = 
0.004 models, suggesting that Z $> 0.004$ in the outer disk.
Small number statistics are an issue for the two inner disk fields, as there are fewer stars 
in these fields with M$_K \approx -8.0$. Based on the presence of stars with M$_K \approx -8$, 
the three outer disk NICI fields contain stars that formed within the past few Gyr, in 
broad agreement with the age distribution of star clusters (e.g. Dong et al. 2008), as well as 
the $K-$band LF of the outer disk fields, which is compared with models in \S 3.3. 
It thus appears that intermediate age stars in the outer regions of M83 are not confined 
to star clusters; rather, these stars are distributed throughout the outer disk. 

\subsubsection{C stars}

	The isochrones shown in Figure 6 assume oxygen-rich atmospheres, and so 
do not track the evolution of stars with carbon-rich atmospheres. 
Cool, bright C stars form in systems with ages $\leq$ a few Gyr (e.g. Maraston 2005), 
and occupy an area of the near-infrared CMDs that extends redward 
of M giants (i.e. evolved stars with oxygen-rich atmospheres). With the caveat 
that comparatively warm C stars may mingle with M giants in near-infrared CMDs, 
cool C stars can be identified based on their near-infrared photometric properties. 

	For this study, C star candidates are identified using the photometric 
properties of C stars in the LMC as a guide. The area of the LMC $(K, H-K)$ CMD that 
contains C stars is marked in Figure 6. The upper and lower boundaries of the C star area, 
as well as the blue limit of the C star sequence, were determined from the envelope of the C 
star locus in the $(K, J-K)$ CMD discussed by Nikolaev \& Weinberg (2000) in their study of 
the LMC. These boundaries were transformed onto the $(K, H-K)$ plane using a 
(J-K) to (H-K) color transformation based on observations of Galactic and LMC/SMC C stars.

	There are a number of sources in the NICI fields with 
photometric properties that are consistent with them being C stars. However, some of these 
are probably background galaxies, the majority of which form a diffuse sequence with $H-K \approx 
0.8$ at these magnitudes (e.g. Davidge 2008). Upper limits on contamination from background 
galaxies can be deduced from the GSC06726--0067 field, where there are three objects 
in the C star regions of the $(K, H-K)$ CMD. If all three are 
background galaxies, then $12 \pm 7$ background objects would be expected in the 
other NICI fields. For comparison, 26 objects fall within the LMC C star regions 
in these four CMDs, and so there is an excess of $14 \pm 9$ objects over that expected 
from presumed background objects. This is a lower limit to the actual number of C stars, as some 
of the red objects near GSC06726--0067 may be C stars that belong to M83. This 
comparison suggests that C stars are present in the fields observed with NICI, and 
such objects are evidence of field stars with an age of a few Gyr in the outer 
regions of M83. 

	The relative numbers of C stars and M giants in a system depends on 
the metallicity and age distributions, and so the ratio of C 
stars to M giants, C/M, is a probe of stellar content. 
The metallicity Z = 0.008 is a benchmark for efforts to model C/M 
as the evolutionary flux measurements for TP-AGB evolution 
can be calibrated from LMC cluster star counts. If stars that formed in the outer 
disk of M83 for the past $\sim 1 - 2$ Gyr did so from material that had a 
metallicity that is no higher than Z = 0.008, then the fuel consumption 
rates compiled by Maraston (2005) predict that all of the fuel consumed during 
TP evolution is done so while the star has an atmosphere with C/O $> 1$; that is, 
all TP-AGB stars that formed between 1 and 2 Gyr should be C stars. Using calibrating 
relations from Maraston (2005), Davidge (2010) concludes that C/M = 2.4, where M is the number 
of thermally-pulsing (TP) AGB M giants, for a Z=0.008 system that has had a constant SFR. 

	The C/M ratio in M83 was measured following the procedures described by Davidge (2010), 
but adjusted for the use of $H-K$ colors. The regions of the CMD in which TP-AGB 
M giants and C stars were counted are indicated in the right hand column of Figure 6. 
TP-AGB M giants are assumed to have $(H-K)_0$ between 0 and 0.45 and M$_K$ between --9.5 and 
$-7 \pm 0.5$. The former is the approximate peak magnitude of the AGB, whereas the latter 
is the approximate point at which evolution on the TP-AGB commences 
for intermediate age stars. Based on the photometric properties of C stars in the LMC, C stars 
are assumed to have $(H-K)_0 \geq 0.40 \pm 0.05$ and M$_K$ between --7 and --9.

	The C/M ratios calculated with the criteria described in the preceeding paragraph 
are listed in Table 2. The uncertainties combine counting statistics with 
the range in star counts that result from the $\pm 0.5$ magnitude 
dispersion in the faint magnitude limit for TP M giants, and the $\pm 0.05$ magnitude 
dispersion in the blue boundary used to identify C stars. The rows 
labelled `Composite' show the C/M ratios for the inner and outer disk groupings.

	Small number statistics render the errors in C/M large. 
Still, with the possible exception of the GSC06726--0067 field, 
the C/M ratio does not vary significantly from field-to-field. This is consistent with 
the majority of the fields having experienced similar star-forming 
histories during intermediate epochs. While this may seem contradictory to the conclusions 
that will be drawn in \S 4, where it is argued that there is a substantial difference in 
stellar content between the inner and outer disk fields, the differences in stellar content 
that are discussed in \S 4 occured during epochs that are not those 
during which substantial numbers of bright C stars are produced.

	The C/M ratios for the composite data sets are lower than predicted 
for a Z=0.008 system that has had a constant SFR. 
Two possible explanations for this difference, involving a metallicity that is higher 
than Z = 0.008 or a star-forming history that is not constant, are considered below.
In reality there will almost certainly be an interplay between both of these, 
and the following discussion is meant to demonstrate the extent to which 
metallicity and star-forming history influence C/M.

	The number of C stars will be lower than predicted if the outer regions of M83 
have Z $> 0.008$, as suggested by the analysis of emission lines in HII regions (Bresolin 
et al. 2009). Maraston (2005) argues that the number of C stars scales inversely with 
metallicity, such that the number of C stars in a system with Z = 0.016 
will be one half that in a Z = 0.008 system, while the number of M giants will 
increase. If this is the case then C/M $\approx 0.7$ 
for solar metallicities, which is much closer to what is measured in the outer disk of M83.

	If the actual SFR has not been constant then this will also influence C/M. A lower SFR $1 
- 2$ Gyr in the past compared with that during the most recent 1 Gyr would result in a lower 
C/M ratio, as the stars that formed within the past 1 Gyr are too massive to have dredged up 
sufficient C into their atmospheres to make them C-rich for a large fraction of their 
TP-AGB lifetimes. While there are as yet no independent 
probes of the SFR in the outer regions of M83 1 -- 2 Gyr in the past, such as very deep CMDs, 
Boissier et al. (2005) find that the global SFRs derived from the slope of the UV-SED and 
from the H$\alpha$ flux are in good agreement, suggesting that the present-day SFR 
is comparable to that averaged over the past Gyr -- unless M83 is being viewed at a 
special epoch then the SFR probably has been constant over 
the past Gyr. If the epoch of constant SFR continued up to 2 Gyr in the past 
then the C/M ratio in M83 is consistent with a solar (or higher) metallicity.

\subsubsection{Red supergiants}

	The outer disk of M83 contains a large number of star clusters with ages $< 100$ Myr 
(Dong et al. 2008); for comparison, the age distribution of clusters in the main body of the 
inner disk peaks near 30 Myr (Jensen, Talbot, \& Dufour 1981). The fraction of red 
supergiants (RSGs) with respect to AGB stars might be expected to grow with R$_{GC}$ 
in the outer disk of M83, given the gradient in UV color found by Thilker et al. (2005). 
Indeed, the UV color measurements made by Thilker et al. (2005) 
suggest that the mean luminosity-weighted age near GSC06726--0067 
(R$_{GC} = 14.5$ kpc) is only a few tens of Myr. It might then be anticipated 
that some of the bright stars in the NICI CMDs are RSGs. 
Of course, the field between clusters may not provide a 
representative census of the youngest RSGs, as the most 
massive stars may not live long enough to diffuse from their place 
of birth, hindering the detection of bright RSGs outside of their natal 
clusters. This is exacerbated by the low density of the outer disk environment, which should 
result in longer diffusion time scales at larger radii when compared with smaller radii (e.g. 
Johnston, Hernquist, \& Bolte 1996). Yet another complication is that the locations of bright AGB 
stars and RSGs overlap on near-infrared CMDs, and only those RSGs with relatively blue colors 
stand out. If they are in dusty star-forming regions, then even young, intrinsically blue RSGs 
may be reddened to the point that they masquerade as AGB stars. 

	Solar metallicity isochrones from Girardi et al. (2002) with ages of 10, 
30, and 100 Myr are compared with the inner and outer disk composite CMDs in Figure 7. 
Both CMD groupings contain objects with $(H-K)_0 < 0.2$ and M$_K < -7$ that are 
candidate RSGs, suggesting that both fields contain stars that formed within the 
past 100 Myr. If, as suggested by the UV colors (Thilker et al. 2005), there is an 
age gradient in the outer disk such that younger luminosity-weighted ages occur at 
larger R$_{GC}$, then a higher fraction of bright RSGs might be 
expected in the GSC06726--0067, 249, and 266 composite CMD (but see the caveat
in the preceeding paragraph regarding diffusion). In fact, the outer disk regions {\it do} 
contain objects with RSG-like colors that are brighter than those in the inner disk field, 
although the statistical significance is modest. Considering the region of the CMDs with M$_K < 
-7.5$ and $H-K$ between 0 and 0.2, there are 5 objects in the outer disk CMD, but none in 
the inner disk CMD. The presence of RSGs notwithstanding, 
the majority of stars near GSC06726--0067, 0259, and 266 have $H-K$ 
colors that are too red for them to be RSGs, indicating that there is 
an underlying substrate of stars with ages $\geq 0.2$ Gyr in these fields. 

\subsubsection{Red giant branch stars}

	The RGB forms a prominent sequence in the CMDs of systems with ages in excess of 
a few Gyr. The 10 Gyr isochrones predict that the RGB-tip occurs near M$_K = 
-6.4$ if Z = 0.004 and M$_K = -6.9$ if Z = 0.019; thus, moderately metal-rich RGB stars 
are well within the detection limits of the NICI data. In fact, there is an increase in the 
number of objects with M$_K < -6$ and $H-K \sim 0.3$ in the composite GSC06726--0126 and 0181 
CMD in Figure 6. Still, the magnitude at which these sources appear is not consistent 
with that expected for the moderately metal-rich RGB that is 
expected in disks (\S 3.3). These faint red objects are not background galaxies, 
as the GSC06726--0067 field, which has the largest R$_{GC}$ and the lowest source density, 
contains only a modest number of sources fainter than M$_K = -6$.

\subsection{$JHK$ Observations of GSC 06726--0126 Field}

	The $J$ observations provide additional information for investigating the near-infrared 
spectral-energy distribution (SED) of sources in the GSC 06726--0126 field. 
The $(K, J-K)$ CMD of the GSC 06726--0126 field is shown in the left hand panel of Figure 8. 
Comparisons are made with Z = 0.019 isochrones from Girardi et al. (2002) in the right hand 
panel of Figure 8. As with the $(K, H-K)$ CMDs, the observed stellar sequence 
cuts diagonally across the isochrones, falling redward of the 1 Gyr isochrone at the 
bright end, and blueward of this sequence at the faint end. 

	The portion of the LMC $(K, J-K)$ CMD from Nikolaev \& Weinberg (2000) that 
contains C stars is indicated in the right hand panel of Figure 8. There are a number of 
sources in this part of the CMD, as expected from the $(K, H-K)$ 
CMD of the same field, and it is likely that the majority of these are C stars. 
Indeed, if the majority of these objects were background galaxies then 
the numbers of objects with $J-K \geq 1.5$ would increase steadily 
towards fainter magnitudes, with no break in counts near M$_K \approx -7$. 
In contrast, the number of objects in Figure 8 with $J-K > 1.5$ dominates when M$_K > -7$. 

	The $(H-K, J-H)$ two color diagram (TCD) provides a 
means of comparing the near-infrared SEDs of sources. The location of 
stars on the near-infrared TCD is sensitive to surface gravity and metallicity, 
in large part because deep $^{12}CO$ and H$_2$O absorption features dominate the near-infrared 
light from stars with surface gravities that are near the lower and upper limits that occur 
in stars. The near-infrared TCD of sources near GSC 06726--0126 is shown in Figure 9. Only 
objects with $K < 22$ are plotted to limit scatter due to photometric errors. 
Fiducial sequences of stars evolving on the main sequence 
and on the AGB from the 1 Gyr Z = 0.019 Girardi et al. (2002) isochrones are also indicated. 
The majority of sources in the TCD have $J-H > 0.7$, which is larger than expected for main 
sequence stars, but is consistent with that expected 
for cool evolved stars with low surface gravity. The model 
AGB sequence passes through the data cloud for sources with $J-K < 1.5$. The locus of LPVs in 
the SMC and LMC, based on data given by Wood et al. (1983; 1985), is also shown, and some 
of the stars in M83 have near-infrared SEDs that are consistent with them being LPVs. 
The comparisons in Figure 9 thus indicate that foreground stars are not present in large numbers, 
and that the sources found here have SEDs that are consistent with those of evolved stars. 

\subsection{The $K$ Luminosity Functions}

	The $K$ luminosity functions (LFs) of objects with $(H-K)_0$ between -0.05 and 0.45 are 
shown in Figure 10. n$_{0.5}$ is the number of sources arcsec$^{-2}$ per 0.5 magnitude 
interval. The turn-down at the faint end of the LFs is due to incompleteness. 
At the bright end there is a general tendency for the number counts to decline towards brighter 
magnitudes, although the LFs of the GSC06726--0126 and 0181 fields are 
relatively flat in the interval between M$_K = -6$ and M$_K = -8$. 

	The mean LFs of the GSC06726--0126 $+$ 0181 and GSC06726--0259 $+$ 0266 fields 
are shown in Figure 11. The CMDs of these fields are also shown to 
allow features in the LFs to be related to those in the CMDs. 
There is modest contamination from background galaxies, and the LFs were corrected for this 
in a statistical manner. A power-law was fit to the GSC 06726--0067 LF in 
the interval between M$_K = -6$ and --9, and the result -- which is shown in the top panel 
of Figure 10 -- was subtracted from the composite LFs. 
Some of the sources near GSC06726--0067 are almost certainly stars belonging to M83, and 
so this procedure overestimates background galaxy contamination. In any event, the 
background galaxy correction has only a minor impact on the LFs shown in Figure 11.

	The LFs of the two field groupings in Figure 11 differ. Whereas the GSC06726--0259$+$0266 
LF follows a single power-law from M$_K = -7.5$ to M$_K = -5.5$, there is a break 
in the number counts at M$_K = -6$ in the GSC06726--0126 $+$ 0181 LF. This change in 
number counts corresponds to a broadening of the CMDs of the GSC06726--0126 and 0181 fields 
that exceeds that expected from photometric errors, and signals the presence of 
stars with a range of ages and/or metallicities (\S 3.1.1). 
The flat nature of the inner disk LFs at the bright end has already been noted, 
and the number counts in the inner disk LFs appear to dip slightly in the interval 
immediately above the break; this behaviour is seen in the individual LFs 
of the GSC06726--0126 and 0181 fields (Figure 10). 

	Model LFs for solar metallicity simple stellar populations were constructed 
to gain insight into the stellar content in these fields, and the models 
are compared with the observed LFs in Figure 12. These models were constructed from 
the Girardi et al. (2002) isochrones using routines in the STARFISH (Harris \& Zaritsky 2001) 
package. The models indicate that the localized dip in number counts above the RGB that occurs in 
the inner disk LFs is a characteristic of systems with 
ages 1 -- 2 Gyr. Indeed, a similar local minimum in star counts 
occurs above the RGB-tip in M32 (Davidge 2000), which is a galaxy that contains a large 
intermediate age population (Davidge \& Jensen 2007, and references therein).

	The inner disk LF appears to have characteristics 
that are indicative of a mix of the log(t$_{yr}$) = 9.0 and older 
LFs. In contrast, the LF of the outer disk is well-matched by the log(t$_{yr}$) = 8.5 LF. The 
model LFs thus suggest that the brightest stars in the inner disk fields tend to be 
older than those in the outer disk fields. 

	If these data contain a substantial old population then the RGB-tip might produce a 
noticeable feature in the LF. To better assess if such a feature is present, 
LFs were calculated with 0.2 magnitude binning, and the results are shown in the top two 
panels of Figure 13. The LFs of the inner and outer disk field are significantly different. The 
GSC06726--0259$+$0266 LF follows a single power-law over the magnitude range shown in 
this figure, and a least squares power-law fit is compared with the LF in Figure 13. 
This power-law was scaled to match the number of points in the GSC06726--0126$+$0181 fields, 
and then subtracted from the composite LF of these fields. The resulting differences between 
the least squares fit and the composite inner disk LF, 
$\Delta$, are shown in the third panel of Figure 13. The $\Delta$ values in three bins differ 
from zero at more than the $2\sigma$ level, and it is clear that 
the power-law that characterizes the GSC06726--0259$+$0266 LF is not 
an adequate representation of the GSC06726--0126$+$0181 LF.

	The composite LFs were convolved with a three-point Sobel edge-detection filter, and the 
Sobel-convolved GSC06726--0126$+$0181 LF is shown in the bottom panel of Figure 13. 
The Sobel-convolved LF of GSC06726--0259$+$0266 shows much smaller bin-to-bin deviations than in 
GSC06726--0126$+$0181, and the dotted lines in Figure 13 indicate the $\pm 1\sigma$ 
variance about the mean convolved signal from that LF. The 
scatter in the Sobel-convolved GSC06726--0259$+$0266 LF provides an 
upper limit to the noise in the edge detection signal. Using this estimate, the break near 
M$_K = -5.8$ in the GSC06726--0126$+$0181 LF is significant at the 3.6$\sigma$ level. 

	Is the change detected in the LF by the Sobel filter due to the RGB-tip? The 
good agreement between the inner disk LF and the log(t) = 9 model in Figure 12 strongly suggests 
that the edge-detection filter is picking up structure in the LF that is due to AGB 
evolution and/or the recent star-forming history. Still, the change in character of the inner 
disk CMDs at the faint end is consistent with the onset of RGB stars (\S 3.1.5). The break 
in the GSC06726--0126$+$0181 LF also has an amplitude of 0.3 -- 0.4 dex, and 
thus is comparable in strength to the RGB-tip discontinuity in the disk of 
NGC 6822 (Davidge 2003), a galaxy that -- like M83 -- 
has formed stars throughout its lifetime (e.g. Tolstoy et al. 2001).

	Perhaps most significantly, the M$_K$ of the break is not consistent with that expected 
from an old moderately metal-rich disk population. Ferraro et al. (2000) define an empirical 
calibration from Galactic globular clusters that relates metallicity and RGB-tip brightness in 
old populations. RGB-tip brightnesses computed from this calibration are indicated at 
the top of Figures 11 and 13. While this calibration is strictly applicable only to old systems, 
the RGB-tip brightness changes only slightly with time for ages in excess 
of a few Gyr, and so it also holds approximately for intermediate age populations.

	If the break in the GSC06726--0126 and 0181 field LFs 
is due to the RGB-tip then the Ferraro et al. (2000) calibration suggests that [M/H] $\sim 
-1.8 \pm 0.1$. The quoted uncertainty reflects the LF binning. This metallicity estimate is that 
of the dominant, presumably `old', population, and is much lower than what would be expected from 
a disk population. It should be noted that if the majority of RGB stars formed within the past 
few Gyr then the Ferraro et al. (2000) calibration predicts an RGB-tip luminosity that is too 
bright, in which case [M/H] inferred from the Ferraro et al. (2000) calibration is a lower 
limit to that of the dominant population. Small number statistics may also 
bias the RGB-tip magnitude to fainter values than would be seen in more populous systems 
(e.g. Crocker \& Rood 1984).

\section{THE SPATIAL DISTRIBUTION OF AGB STARS: EVIDENCE FOR AN AGE GRADIENT IN THE OUTER DISK}

\subsection{Radial Trends in the Star Counts}

	Disk light profiles can be sorted into three broad types, based on their 
behaviour at large radii (e.g. Pohlen \& Trujillo 2006). Type I disks 
have light profiles that follow a single exponential. While the classical exponential 
profile was once thought to be the norm for disks, 
with any break in the light profile at large radius associated with 
the physical truncation of the stellar disk (e.g. van der Kruit \& Searle 1981), recent surveys 
have found that only $\sim 10\%$ of spiral galaxies have Type I profiles. Type II and III 
disks have profiles that break either down (`truncated') or up 
(`anti-truncated') at large radii, respectively.

	An important caveat when interpreting this 
classification scheme is that light is only a crude proxy for mass. 
Bakos, Trujillo, \& Pohlen (2008) suggest that the 
break in Type II light profiles may be due to a change in stellar 
content induced by the physics of star formation in low density environments, rather than a 
structural change in disk properties. They conclude that, despite having different 
light profiles, the mass profiles of Type I and II disks may be similar. The similar 
kinematic properties of PNe in the outer regions of M83 and M94, which are galaxies that 
have very different light profile characteristics at large radii (Herrmann et al. 2009), 
is consistent with this.

	Resolved stars provide a more direct probe of stellar content than integrated light. 
The spatial distribution of stars with $K \leq 22.5$ in the NICI fields is 
investigated in Figure 14, where projected stellar density is shown as a 
function of R$_{GC}$. The error bars show counting statistics. With the exception of the 
area around GSC06726--0067, the density of AGB stars is more-or-less constant 
between de-projected radii of 350 and 500 arcsec, with $\approx 0.2$ 
stars with $K < 22.5$ arcsec$^{-2}$. This is counter to the behaviour of 
the near-infrared light profiles of M83 measured by Jarrett et al. (2003), which break 
downward at R$_{GC} = 360$ arcsec. Thus, the distribution of bright red stars in the outer 
disk of M83 does not track the integrated near-infrared light.

	The specific frequency (SF) measures the number of objects per unit integrated light. 
In the context of a simple stellar systems with ages $\geq 0.2$ Gyr, the SF of AGB stars 
decreases with increasing age (e.g. Maraston 2005). Thus, higher AGB SFs correspond to 
younger luminosity-weighted ages. 

	In the current study, the SF is defined as the number of stars with $K < 22.5$ per M$_K 
= -16$ brightness, with the latter calculated from the Garrett et al. (2003) 2MASS $K-$band light 
profile. The SF measurements for the GSC06726--0126, 0181, 0259, and 0266 fields are shown in 
Table 3; GSC06726--0067 is too far from the center of M83 to have reliable 2MASS surface 
photometry, and so a SF was not computed for this field. The light 
profile was extrapolated by $\sim 30$ arcsec to compute the SF of 
the GSC06726--00259 field. There are field-to-field differences in the SFs of AGB stars; 
whereas the SF of AGB stars near GSC06726--0181 and 0266 are very similar, 
the SF of AGB stars around GSC06726--0126 is comparatively low, while the SF of AGB stars 
near GSC06726--00259 is very high. 

	If the stellar content throughout the main body of M83 is similar to that in 
the outer disk then the light profile of M83 can be transformed into an AGB 
number density profile by applying a scaling factor determined from calibration 
fields. For this study, the GSC06726--0126, 0181, and 266 fields, 
which are at radii where the $K-$band surface brightnesses are most secure, were used to set 
the calibration. The resulting AGB density profile predicted from this calibration 
is shown as the solid line in Figure 14.

	A linear extrapolation of the solid line in Figure 14 to larger R$_{GC}$ falls 
well below the AGB star densities measured near GSC06726--0259 and 0067. To the extent that 
light profiles can be extrapolated to larger radii then this suggests that the fraction of the 
total $K-$band light that originates from bright AGB stars changes with radius in the outer 
disk of M83, in the sense of increasing towards larger radii; 
the luminosity-weighted age thus drops (i.e. becomes younger) towards larger radii. 
This is consistent with the trend of bluer UV colors towards larger radii measured in 
M83 by Thilker et al. (2005), and the absence of an RGB-tip feature in the LF of 
stars near GSC06726--0259 and 0266. A similar trend of decreasing 
luminosity-weighted age with increasing radius is seen in the outer disk of M81 (Davidge 2009).

\subsection{Star Counts as Probes of Stellar Content}

	The fraction of the $K-$band light that comes from AGB stars 
is a probe of stellar content. Based on the SFs listed in Table 3, 
AGB stars in the inner disk fields contribute on average only $\sim 20\%$ of the 
$K-$band light that comes from AGB stars in the outer disk fields. 
This can be used to estimate the difference in age between the inner and 
outer disk fields using the relation between the fraction of total $K-$band light 
from AGB stars and age computed by Maraston (2005) for simple stellar populations. 
The solar metallicity relation in Figure 13 of Maraston (2005) indicates that a difference in the 
contribution made by AGB stars like that between the inner and outer disk fields 
corresponds to an age difference of $\sim 0.6$ dex, or a factor of 4. If the 
luminosity-weighted age of the inner disk fields is a few Gyr, 
then the Maraston (2005) calibrations predict that RGB stars should contribute $\sim 6\times$ 
less light in the outer disk fields than in the inner disk fields. Thus, one would 
expect a much weaker RGB-tip feature in the outer disk LF than in the inner disk LF.
 
	The SF measurements can be used to compare the stellar content of 
the outer disk of M83 with that of other disk systems in a purely empirical manner. 
Although it has a moderately low SFR at the present day, 
M31 is an interesting comparison object since there was more vigorous 
star-forming activity $\approx 1$ Gyr in the past (e.g. Williams 2002). The structural 
properties of M31 also suggest that it is a `typical' spiral galaxy in terms of the cumulative 
impact of interactions on its angular momentum content (Hammer et al. 2007). 

	The SF of stars in the M31 Disk 2 field, where the density of AGB stars with M$_K 
\leq -6$ is 0.50 arcsec$^{-2}$ and the $K-$band surface brightness is 17 mag arcsec$^{-2}$ 
(Davidge et al. 2005; Olsen et al. 2006), is shown in Table 3. The SF 
of AGB stars in M31 Disk 2 falls midway between the extremes in the M83 measurements. This is 
consistent with the three NICI fields with the smallest R$_{GC}$ having luminosity-weighted 
ages that bracket the luminosity-weighted age of the M31 disk. This does not mean necessarily that 
M83 and M31 have had similar star forming histories; rather, the star-forming histories of both 
galaxies during intermediate epochs has lead to them having similar cumulative SFs. 

\section{SUMMARY AND DISCUSSION}

	Deep near-infrared images recorded with NICI on GS
have been used to investigate the spatial distribution 
and photometric properties of stars in the outer regions of the nearby barred spiral galaxy 
M83. Five fields that together sample a total area of $\sim 10^3$ arcsec$^{2}$ (0.6 kpc$^2$ 
at the distance of M83), were observed. The fields are centered on bright NGSs, and 
sample large galactocentric radii. The GSC06726--0126 and 
0181 fields contain a mix of stars from the classical disk and outer 
disk of M83, while the stellar content of the outer disk is charted in a more 
pristine way with the GSC06726--0067, 0259, and 0266 fields. 

	Studies of the resolved stellar content of a system provide a direct check of 
conclusions drawn from the analysis of integrated light. A general result of the current 
investigation is that the ages and spatial distributions of the brightest resolved 
stars in the outer disk of M83 are consistent with the properties of the integrated UV light. 
First, the presence of bright AGB stars in all of the fields indicates that the outer disk of 
M83 contains a wide-spread stellar component that formed $\geq 0.2$ Gyr in the past, in agreement 
with the luminosity-weighted age inferred from the integrated UV colors of the diffuse light 
around M83 and of star clusters in the outer disk. It is further worth noting that roughly half 
of the clusters investigated by Dong et al. (2008) using integrated IR light in the outer disk of 
M83 have ages $> 0.2$ Gyr, with the oldest having ages $> 1$ Gyr.
The presence of AGB stars in the NICI data that span a wide range of photometric properties 
is consistent with such an age dispersion in the outer regions of M83. 

	Second, the NICI data indicate that there is an age gradient in the outer regions of M83. 
Based on a comparison of LFs with models, the AGB stars in the GSC06726--0126 and 0181 fields 
are older on average than in the fields with larger R$_{GC}$. The FUV -- NUV 
color in the outer regions of M83 varies with radius (Thilker et al. 2005), in a manner that 
is consistent with an age gradient similar to that measured from the AGB stars. 

	Third, the number density of stars with $K < 22.5$ (M$_K > -6$) 
is constant with radius between R$_{GC} =$ 10 and 13 kpc, with a projected 
density of 0.2 stars arcsec$^{-2}$ (340 stars kpc$^{-2}$). The density of AGB stars 
in the GSC06726--0067 field, which is at R$_{GC} = 14.5$ kpc, is markedly 
lower than at smaller radii. The flat density distribution of 
individual stars at intermediate radii followed by a drop at larger radii 
is consistent with the integrated UV light, where the light profile flattens near 
10 kpc before dropping at larger radii (Figure 5 of Thilker et al. 2005).

	The agreement between the properties of resolved stars and integrated UV light is 
gratifying, as the UV light from intermediate age systems originates predominantly from 
main sequence stars with masses $\geq 2$M$_{\odot}$, which have a lifetime $\leq 1$ Gyr. 
The stars that are traced by the UV light thus formed at the same 
time as the (only very slightly) more massive progenitors of the brightest AGB stars. An 
obvious avenue for future work is to measure the density of bright AGB stars over a 
larger fraction of the outer regions of M83. If, as suggested by the NICI data, bright 
AGB stars follow the UV light, then the number density of bright AGB stars in the field should 
drop dramatically at major axis radii in excess of 8.5 arcmin.

	This investigation highlights the capabilities of NICI for deep photometric studies, and 
provides insights into the site characteristics of Cerro Pachon that relate to AO. 
While the design of NICI is optimized to address the challenges encountered in high contrast 
imaging, the observations presented here indicate that NICI can also be employed 
for deep photometric studies of faint objects over angular scales of $\approx 10$ arcsec from the 
guide star. The angular resolution of these data approaches the diffraction limit 
of the telescope. While modest signatures of anisoplanaticity are seen, 
the impact on the photometric measurements is not large, and can be compensated for 
by applying a spatially variable PSF. In fact, for three of the five 
fields the PSF was sufficiently stable across the $14 \times 14$ arcsec$^2$ field that a 
spatially variable PSF was not required. 

\subsection{AGB Stars as Probes of the Recent Star Formation Rate}

	One condition for triggering star formation is that the density of cool gas be high 
enough to enable gravitational collapse. Observations indicate that the threshold gas density for 
collapse is $\sim 0.7 \times$ the Toomre stability statistic (Kennicutt 
1989; Martin \& Kennicutt 2001). While environments with gas densities 
that are typically lower than the critical value, such as the outermost regions of disks, are not 
expected to be sites of large-scale star formation, localized pockets of material with 
densities that are sufficient to trigger star formation may still occur. Accounting for 
such pockets in simulations of low density environments results in star-forming histories that 
are in better agreement with observations than models that assume a homogeneous ISM 
(e.g. Orban et al. 2008).

	The occurence of star-forming pockets in the peripheral regions of disks 
is evidently not a rare phenomenon, as star clusters with 
ages $< 400$ Myr are seen at large radii in many nearby 
galaxies. Such clusters are typically traced out to 2R$_{25}$, leading Zaritsky \& 
Christlein (2007) to conclude that star formation in the outer regions of disks is `common and 
long lasting'. Welikala et al. (2008) investigate the distribution of star-forming activity 
throughout galaxies in a range of environments. While star formation is most intense in the 
inner regions of galaxies, limited levels of star formation are consistently 
traced out to large radii, and there is only a modest radial decline in star-forming 
activity at large radii. 

	The average SSFR computed from UV emission averaged over the entirety of the outer 
regions of M83 is $1 \times 10^{-5}$ M$_{\odot}$ year$^{-1}$ kpc$^{-2}$ (Bresolin et al. 2009). 
The five fields observed with NICI cover $10^3$ arcsec$^2$, or $\approx 0.65$ kpc$^2$. 
Therefore, if the SSFR has been fixed at $1 \times 10^{-5}$ M$_\odot$ kpc$^{-2}$ then $8 \times 
10^3$ M$_{\odot}$ of stars would have formed between 200 Myr and 2 Gyr in the past. Adopting a 
mass-to-light ratio M/L$_{bol} \sim 0.3$ (Mouhcine \& Lancon 2002a), then the total luminosity 
of stars formed in the past 2 Gyr is M$_{bol} \sim -6.3$. Fuel consumption values from Maraston 
(2005) predict that TP-AGB stars should contribute $\sim 20\%$ of the 
light, or a total M$_{bol} \sim -4.6$, to a system with an age of 
2 Gyr. Given that TP-AGB stars have M$_{bol} \approx -5$ (Mouhcine \& Lancon 2002b), 
which translates into M$_K \geq -7$, then the Bresolin et al. (2009) SSFR predicts at most 
$\approx 1$ TP-AGB star in the five NICI fields. Given that there are many more bright AGB stars 
than this then the SSFR during intermediate epochs 
{\it in the south eastern quadrant of the outer disk of M83} 
must have been higher than the Bresolin et al. (2009) estimate. 
In other words, the south east quadrant of M83 appears to have been a region with 
a localized excess in star-forming activity for the past $\approx$ Gyr 
when compared with the rest of the outer disk. The Bresolin et al. 
(2009) SSFR suggests that the density of bright AGB stars in other regions of the M83 
outer disk should be markedly lower than in the south east quadrant, and this can be 
checked with observations of other fields throughout the outer disk of M83. 

	Using the integrated $8\mu$m flux from star clusters, Dong et al. (2008) estimate that 
the SSFR in the outer disk of M83 has been $8.0 \times 10^{-4}$ M$_{\odot}$ year$^{-1}$ 
kpc$^{-2}$. The clusters studied by Dong et al. (2008) are in the southern portion of the 
outer disk, with a projected distance of only a few kpc from the fields observed with NICI.
This SSFR is an upper limit when applied to intercluster regions for 
two reasons. First, and most obviously, star clusters have stellar 
densities that exceed those in the field and may contain stars with lifetimes that 
are too short to allow them to disperse from their birthplaces before 
ending their evolution. Second, as Dong et al. (2008) note, 
emission from PAHs, which are not tracers of star-forming activity but will 
still contribute to the flux at 8$\mu$m, will skew the SSFR to higher values. 

	The Dong et al. (2008) SSFR predicts that 
the total luminosity of stars that would have formed in the past 2 Gyr 
is M$_{bol} \approx -11.1$, and so M$_{bol} \approx -9.3$ would originate from TP-AGB stars. 
There would then be $\approx 50$ stars with M$_{bol} = -5$ in all five NICI fields, whereas 
100 TP-AGB stars are present. On first blush, it may appear that the Dong et al. (2008) 
SSFR is too low, which is contrary to the expectation that it might overestimate the SSFR. 
Still, if the AGB star counts near GSC06726--0126 and 0181 are excluded because of their 
proximity to the classical disk of M83, then the Dong et al. SSFR predicts that 
there should be 30 TP-AGB stars in the remaining fields, whereas 33 AGB stars are seen. 
There is thus good agreement with the numbers of bright AGB stars in the 
fields investigated with NICI if the fields that are least prone to contamination from 
the classical disk of M83 are considered. These numbers suggest that the SSFR 
estimated from clusters in the southernmost regions of M83 also applies to the nearby field. 
Such agreement also suggests that the characteristic timescale for cluster disruption in the 
outer disk of M83 is much lower than $\sim 1$ Gyr; otherwise, stars would stay locked in clusters 
and the SSFR inferred from the field would be much lower than observed.

\subsection{Possible Origins of the Gas and Stars in the Outer M83 Disk}

	Bush  et al. (2008) model the evolution of a disk that has an extended, constant 
density gas component. Spiral density waves create areas of high density in the outer disk 
that are sites of star formation. A diffuse stellar component subsequently 
forms as areas of active star formation propogate with azimuth. If an extended gas disk 
like that modelled by Bush et al. (2008) has been in place around M83 for a significant 
fraction of the Hubble time then an old, diffusely distributed stellar component 
will have formed {\it in situ}. That a substantial diffuse RGB component has not been 
detected in the three outermost fields argues that the outer gas disk around M83 has not 
been in place for longer than a few Gyr.

	The processes that may cause the formation of structurally distinct inner 
and outer disk components may leave signatures in the stellar contents of disks. 
A downward breaking profile, such as is seen in the near-infrared integrated light 
profiles of M83, may be due to the reduced frequency at which the star formation threshold 
is breeched in the outer regions of gas disks with a radially decreasing density profile 
(Elmegreen \& Hunter 2006). Roskar et al. (2008) model the impact of star formation and 
dynamical processes on disk structure, and find that systems with a 
downward breaking stellar mass profile have an age minimum at the break location, due in 
part to the decrease in gas density towards larger radii. The 
mean age of stars in the outer disks of these models increases towards larger radii 
due to the orbital heating of old stars by secular processes. 

	Stars in disks may migrate over substantial radial distances. 
Sellwood \& Binney (2002) find that stars that formed at 
intermediate radii in disks may be perturbed by spiral waves and move out to 
more than twice their radius of formation. The bombardment 
of the outer disk of M83 by halo sub-structures, which is a process that has been 
suggested as a means of explaining the kinematic characteristics of PNe in M83 
(Herrmann et al. 2009), may also contribute to mixing in the M83 disk.
The flat radial distribution of bright AGB stars in the four innermost NICI fields suggests 
that they were not scattered to large radii from the inner regions of the disk by 
interactions with spiral structure. 

	The plateau in AGB number counts that is defined by all of the NICI fields 
save that around GSC06726--0067 is reminiscent of an anti-truncated light profile. 
Such profiles may be the result of the accretion of material with an angular 
momentum distribution that is systematically different from that of the main body of the disk 
(van der Kruit 2007), dynamical evolution spurred by galaxy-galaxy interactions (e.g. Younger 
et al. 2007), and/or dynamical processes that stir the outer regions of disks (de Jong et al. 
2007). Younger et al. (2008) find that the re-distribution of gas and stars resulting from the 
transfer of angular momentum in interactions may produce anti-truncated light profiles. 
Simulations discussed by Younger et al. (2008) predict that older 
stars are re-distributed to larger radii, in response 
to the transfer of angular momentum as gas is channeled inwards. The stellar content of the 
outer disk is thus skewed to an older luminosity-weighted age than 
was in place prior to the interaction. Of course, not all of the gas in a disk may be 
channeled into the galaxy center, and if some gas is moved to 
large radii by tidal forces, or is stripped from one galaxy and settles in the 
disk plane of the other then it may eventually be an area of star formation, 
thereby producing an extended young stellar component, such as is seen in M83.

	Erwin, Beckman, \& Pohlen (2005) find that some galaxies with anti-truncated light 
profiles show signatures of tidal interactions, suggesting that 
galaxy-galaxy interactions may be related to the formation of anti-truncated systems. 
It has been suggested that the elevated star forming activity in M83 may have 
been triggered by an interaction with NGC 5253 (van den Bergh 1980, but see also Rogstad et 
al. 1974), or KK 208 (Karachentsev et al. 2002). Bresolin et al. (2009) conclude that a tidal 
origin for the outer disk of M83 can not be ruled out. 

	There are indications that M83 has been subjected to tidal 
interactions. The HI distribution in M83 is asymmetric, with $1.4 \times$ more HI on the 
eastern side of the galaxy than the western side (Huchtmeier \& Bohnenstegel 1981). The outer 
gas disk is also warped (e.g. Rogstad et al. 1974), a phenomenon that has been attributed to 
interactions with satellites in other systems (e.g. Hunter \& Toomre 1969). Diaz et al. (2006) 
find a mass concentration in the central regions of M83 that may be the remnants of a companion 
that was accreted by M83. The kinematics of HI clouds seen in the vicinity of M83 are 
also consistent with a possible tidal origin (Miller et al. 2009). The absence of a 
metallicity gradient in the outer disk of M83 (Bresolin et al. 2009) suggests that this 
material is well-mixed, as might be expected if it has been churned by tidal forces, although 
Bresolin et al. (2009) point out that a homogeneous metallicity distribution may simply be a 
consequence of the low surface density of gas, much like that seen in dwarf galaxies. 
Models of disks that are subject to the infall of material from the surrounding halo can 
also produce flat abundance profiles at large radii (e.g. Magrini et al. 2009). 
An analysis of the orbital properties of the galaxies near M83 will help to determine 
when M83 was last subjected to a significant interaction, and if this event can be associated 
with the epoch of elevated star formation in its outer disk.

\acknowledgements{It is a pleasure to thank the NICI support scientist, Thomas Hayward, for 
invalueable advice and assistance in defining the observations and recording these data. Thanks 
are also extended to the anonymous referee, who made suggestions that improved the final paper.}

\clearpage

\begin{table*}
\begin{center}
\begin{tabular}{lcccr}
\tableline\tableline
GSC \# & RA & Dec & $R$ & R$_{GC}$ \\
 & (2000) & (2000) & (mag) & (kpc) \\
06726--00126 & 13 37 17.8 & --29 55 05.8 & 13.3 & 8.8 \\
06726--00181 & 13 37 14.7 & --29 56 17.6 & 10.5 & 9.7 \\
06726--00259 & 13 37 10.3 & --29 58 10.7 & 11.5 & 11.9 \\
06726--00266 & 13 36 51.5 & --29 58 21.6 & 10.7 & 10.5 \\
06726--00067 & 13 37 38.4 & --29 53 33.4 & 11.0 & 14.5 \\
\tableline
\tableline
\end{tabular}
\end{center}
\caption{Guide Star Locations and $R$ Magnitudes}
\end{table*}

\clearpage

\begin{table*}
\begin{center}
\begin{tabular}{lc}
\tableline\tableline
GSC \# & C/M \\
 & \\
06726--00067 & $1.2 \pm 0.6$ \\
06726--00126 & $0.3 \pm 0.2$ \\
06726--00181 & $0.6 \pm 0.6$ \\
06726--00259 & $0.5 \pm 0.5$ \\
06726--00266 & $0.3 \pm 0.2$ \\
Composite 0126, 0181 & $0.4 \pm 0.3$ \\
Composite 0067, 0259, 0266 & $0.5 \pm 0.3$ \\
\tableline
\tableline
\end{tabular}
\end{center}
\caption{C/M Ratios}
\end{table*}

\clearpage

\begin{table*}
\begin{center}
\begin{tabular}{lc}
\tableline\tableline
Field \# & SF \\
 & \\
06726--00126 & $860 \pm 90$ \\
06726--00181 & $2800 \pm 300$ \\
06726--00259 & $15600 \pm 2000$ \\
06726--00266 & $3000 \pm 400$ \\
M31 D2 & $1250 \pm 100$ \\
\tableline
\tableline
\end{tabular}
\end{center}
\caption{Specific Frequencies of Stars with M$_K \leq -6$}
\end{table*}

\clearpage

\begin{figure}
\figurenum{1}
\epsscale{0.95}
\plotone{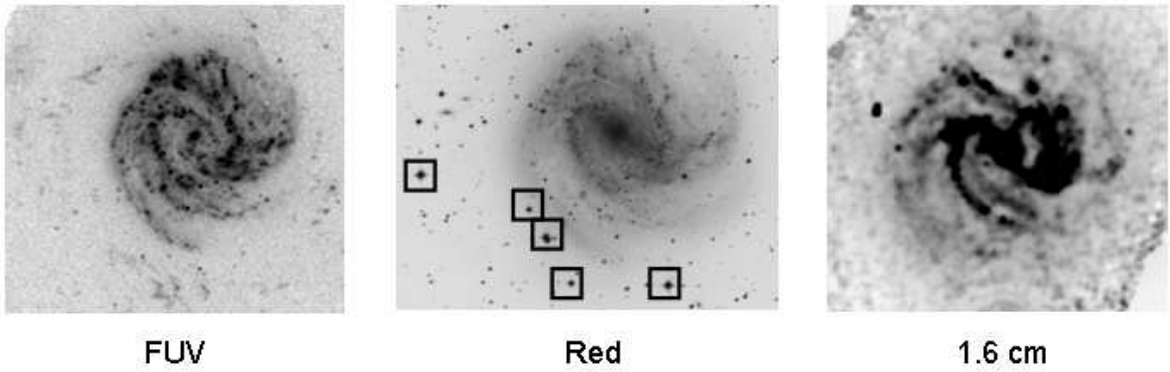}
\caption{Images of M83 at three different wavelengths. The FUV GALEX 
image from Thilker et al. (2005), the red Digital Sky Survey image, and the 1.6 cm map from 
Tilanus \& Allen (1993) are shown; north is at the top, and east is to the left. The images 
are shown with a common angular scale to facilitate comparisons, and the NICI fields 
are indicated on the DSS image. The actual area observed around each NGS is much smaller 
than the boxes used to mark the NGSs.}
\end{figure}

\clearpage

\begin{figure}
\figurenum{2}
\epsscale{0.75}
\plotone{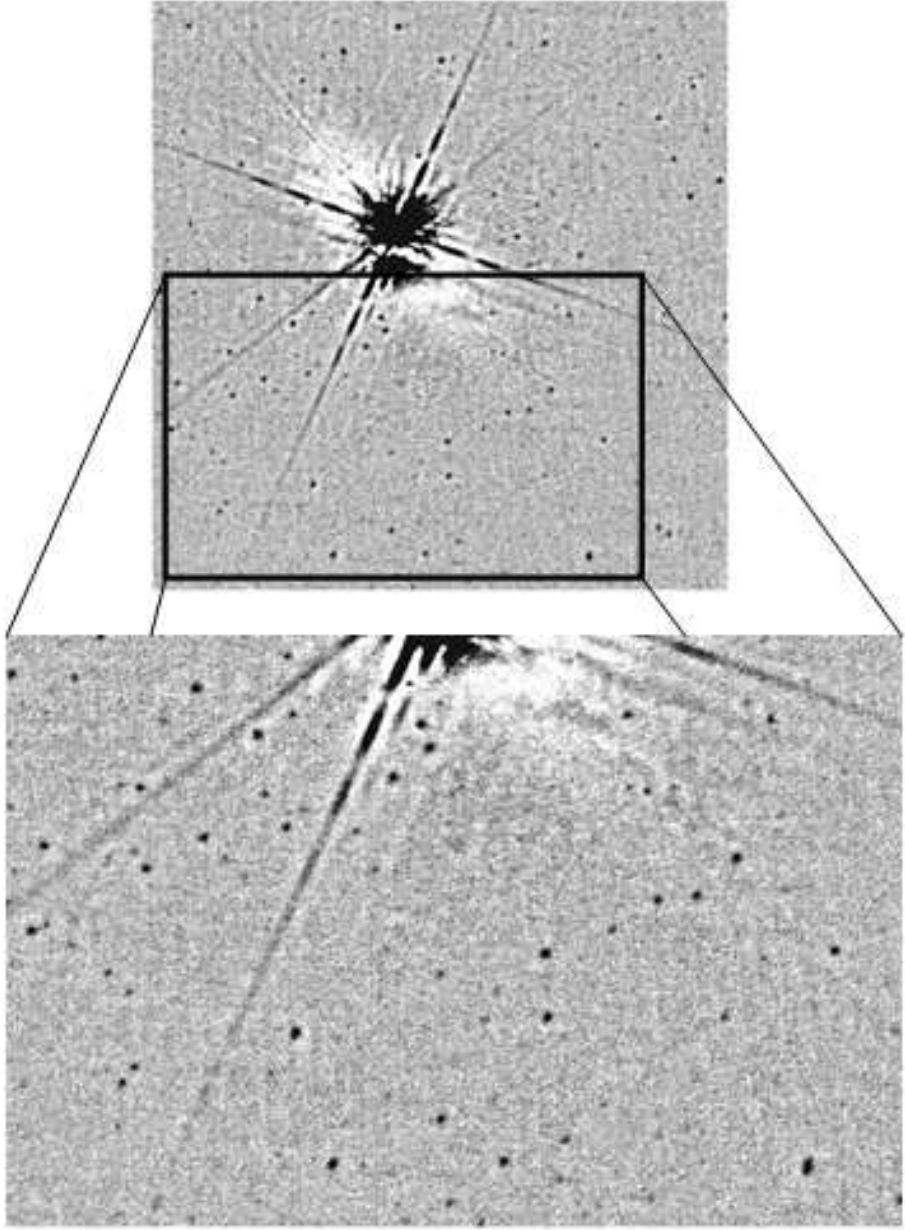}
\caption{The processed $H-$band image of GSC 06726-00259. North is at the top, and east is to the 
left. The image in the top panel covers $14 \times 14$ arcsec$^2$. 
Speckle noise hinders the detection of faint sources 
near the NGS. The lower panel shows an expanded view of part of the field. 
Numerous sources are seen, the majority of which are AGB stars in the 
outer disk of M83.}
\end{figure}

\clearpage

\begin{figure}
\figurenum{3}
\epsscale{0.75}
\plotone{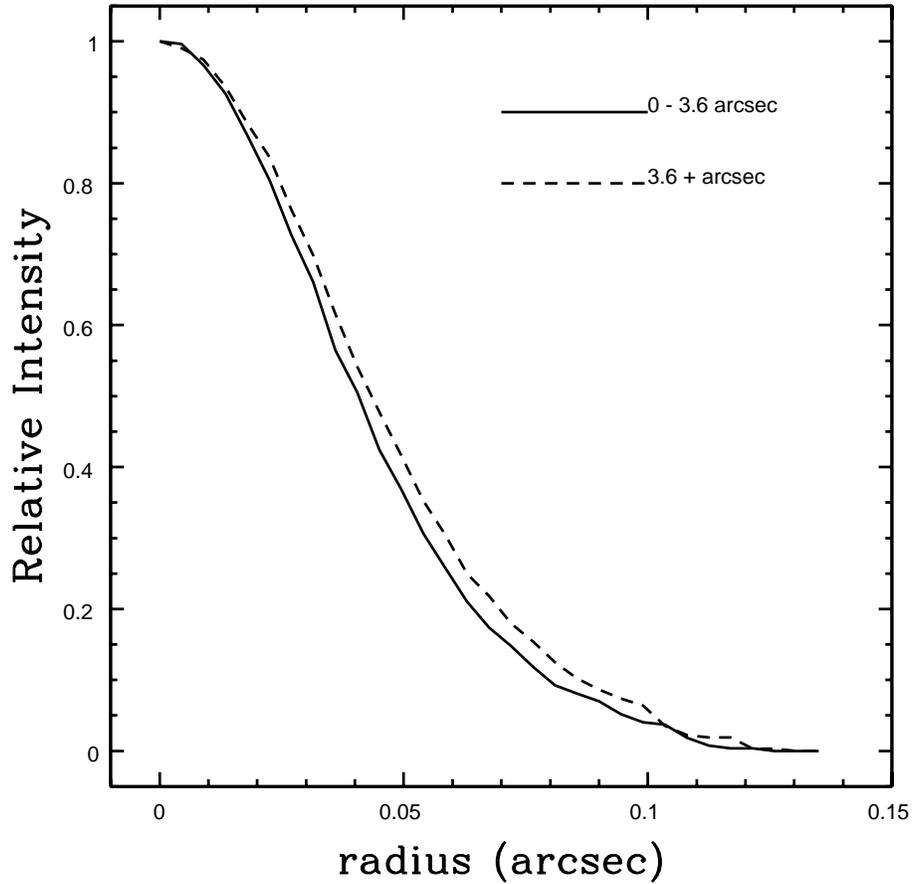}
\caption{Azimuthally averaged $H-$band PSF profiles as constructed from sources with 
$r_{NGS} < 3.6$ arcsec (solid line), and $r_{NGS} > 3.6$ arcsec (dashed 
line), where $r_{NGS}$ is the angular offset from GSC06726--00259. 
Both PSFs have been normalized to their peak intensities. 
The FWHMs of the two PSFs are 0.080 arcsec and 0.086 arcsec.} 
\end{figure}

\clearpage

\begin{figure}
\figurenum{4}
\epsscale{0.75}
\plotone{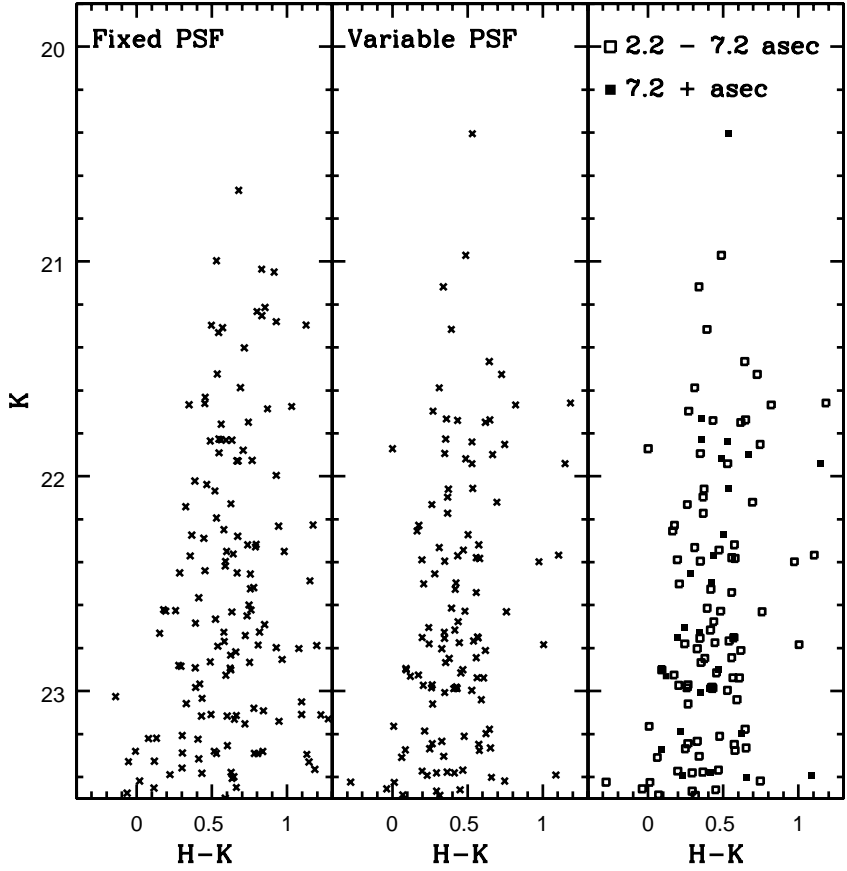}
\caption{$(K, H-K)$ CMDs of the GSC06726--0259 field that were obtained by applying fixed 
(left hand panel) and spatially variable (middle panel) PSFs. 
The photometric measurements made with a variable PSF results 
in a tighter sequence than those made with a fixed PSF. 
This is one of only two fields (the other is GSC06726--0266) in which a variable PSF 
yields a tighter photometric sequence than a fixed PSF. The right hand panel shows the 
data plotted in the middle panel, but with the points identified based on $r_{NGS}$. 
Neither the ridgelines or scatter envelopes of the sequences defined by stars in the 
two $r_{NGS}$ ranges differ, indicating that spatial variations in the PSF do not contribute 
substantial systematic photometric errors. This indicates that DAOPHOT has tracked PSF 
variations to a level where residual errors in the variations are modest when compared 
with those due to random noise (\S 3.1).}
\end{figure}

\clearpage

\begin{figure}
\figurenum{5}
\epsscale{0.75}
\plotone{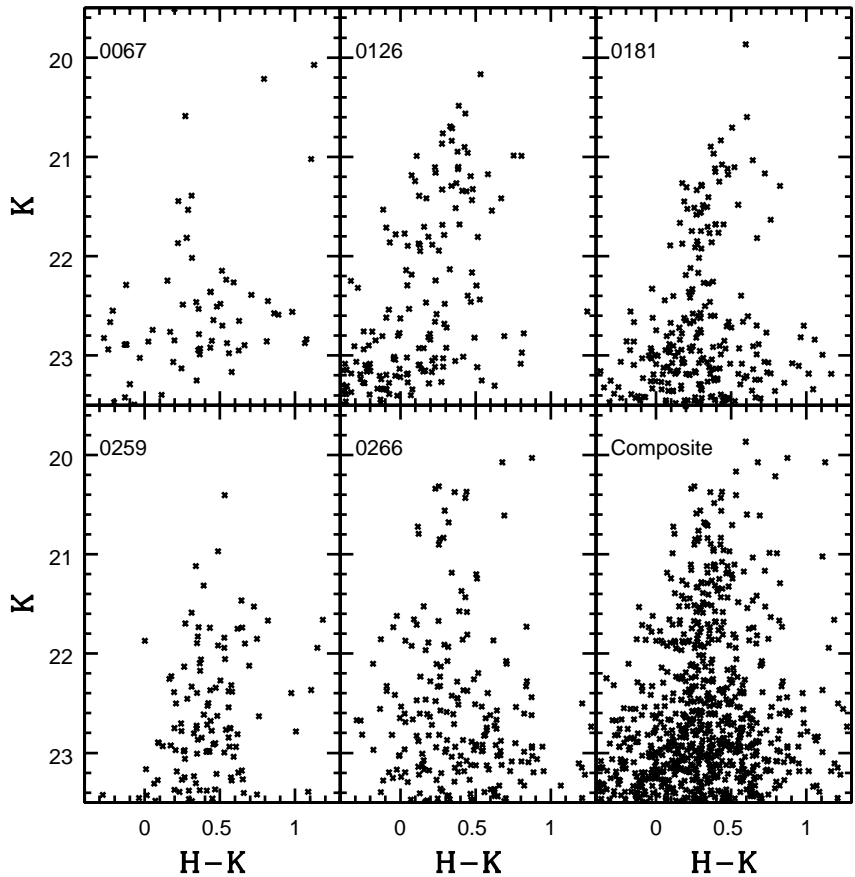}
\caption{The $(K, H-K)$ CMDs of the individual fields, and the composite CMD 
of all fields. The bright sequence centered near $H-K \approx 
0.4$ is dominated by AGB stars, although some RSGs are probably also 
present (\S 3.1.4). If an old moderately metal-rich population is present then the RGB-tip 
will occur near $K \approx 22$. Note that the GSC06726--0126 
and 0181 CMDs differ from those of the others near the bright end, in that the AGB 
sequence departs from vertical. Because they are viewed close to the south east spiral arm, 
these fields are expected to have the greatest amount of 
contamination from stars in the classical disk of M83.}
\end{figure}

\clearpage

\begin{figure}
\figurenum{6}
\epsscale{0.75}
\plotone{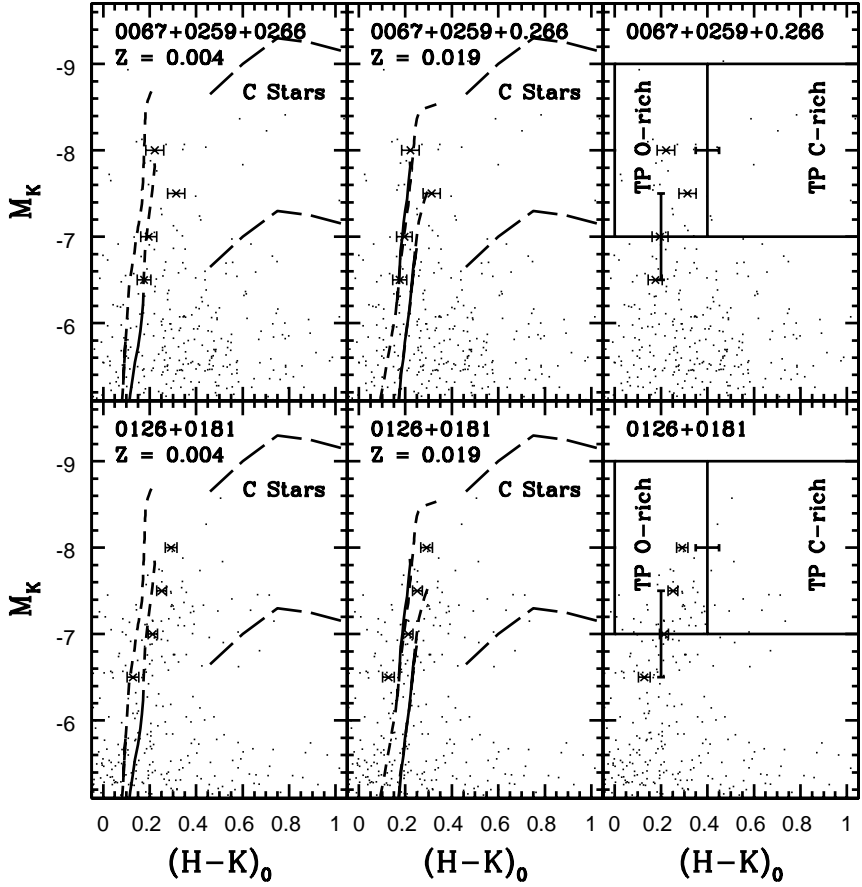}
\caption{Composite CMDs of the NICI fields are compared with 1 Gyr and 10 Gyr isochrones 
from Girardi et al. (2002). The left hand isochrone in each panel is the 1 Gyr sequence. 
A distance modulus of 28.6 (Karachentsev et al. 2007) and an 
extinction A$_V = 1.4$ (\S 3.1.2) have been assumed. 
The portions of the isochrones that cover evolution on the RGB are shown as solid 
lines, while the evolution on the AGB above the RGB-tip is indicated with short dashed lines. The 
long dashed lines mark the area of the LMC CMD that contains cool C stars, based on 
the C star sequence in the Nikolaev \& Weinberg (2002) CMD. The mean colors in $\pm 0.25$ 
magnitude intervals in $K$ for stars with $(H-K)_0 \leq 0.45$ are also shown for each 
composite CMD, and the error bars show the $1\sigma$ uncertainties in these means. 
The stars with M$_K \geq -7.8$ have intermediate ages, and the mean $H-K$ 
colors of both field groupings better match the solar metallicity sequences at M$_K = -8$ 
than the sub-solar metallicity sequences. The areas of the CMDs used to count 
M giants and C stars are indicated in the right hand panel. 
The error bars in this panel show (1) the uncertainty in M$_K$ 
that corresponds to the onset of TP evolution for M giants, and (2) the uncertainty in the 
$H-K$ color that separates the areas of the CMD that are dominated by M giants and C stars.
at M$_K = -8$.}
\end{figure}

\clearpage

\begin{figure}
\figurenum{7}
\epsscale{0.75}
\plotone{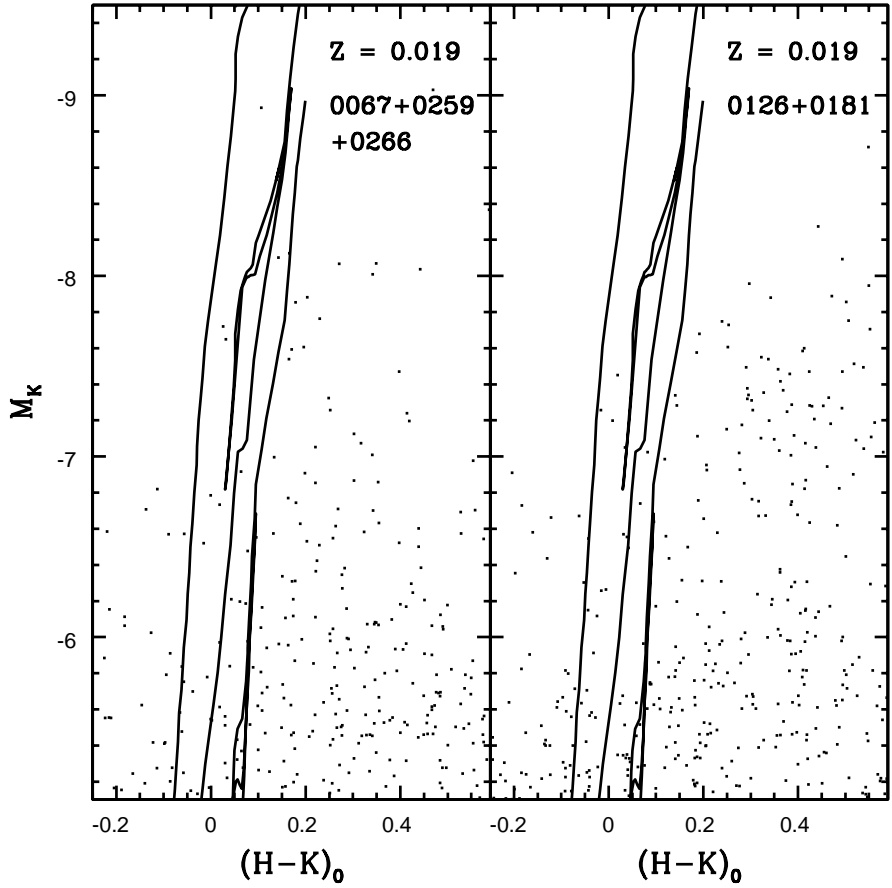}
\caption{The same as Figure 6, but showing Z = 0.019 isochrones with ages 10 Myr, 30 Myr, and 
100 Myr. Both the inner and outer disk field groupings contain stars with $(H-K)_0 < 0.2$ 
and M$_K < -7$ that, based on their colors, are candidate RSGs.}
\end{figure}

\clearpage

\begin{figure}
\figurenum{8}
\epsscale{0.75}
\plotone{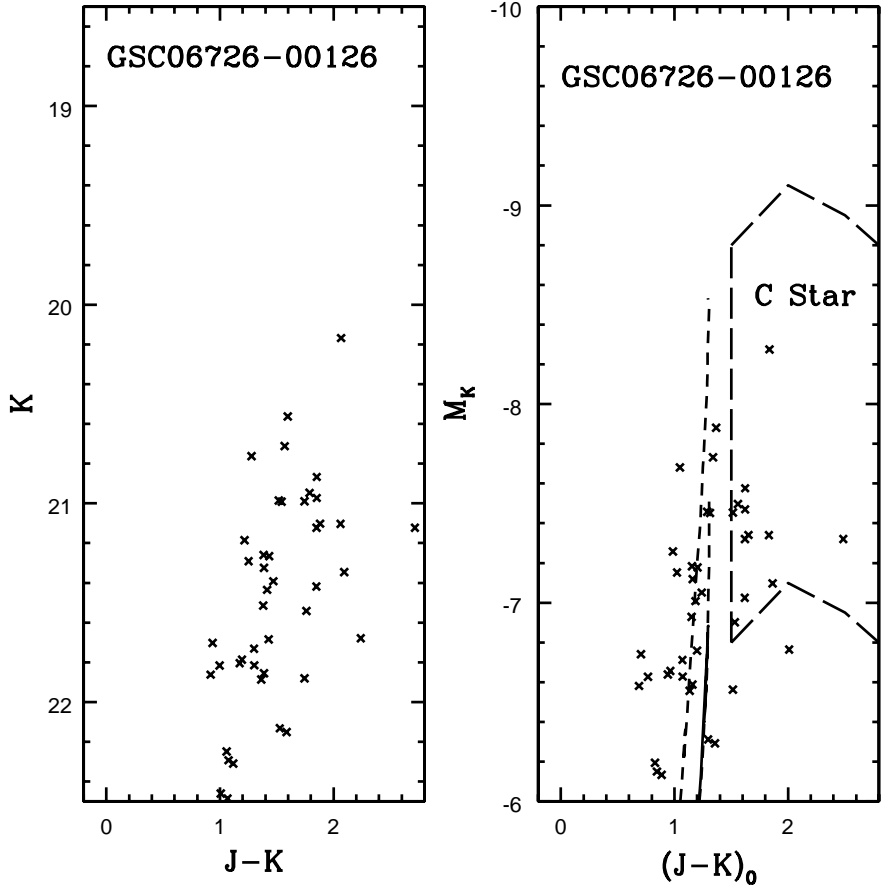}
\caption{The $(K, J-K)$ and $(M_K, J-K)$ CMDs of objects near GSC06726--00126. 
Isochrones from Girardi et al. (2002) with ages 1 and 10 Gyr and Z=0.019 
are shown in the right hand panel, with evolution on the RGB 
(solid line) and AGB (dashed line) indicated. A distance modulus of 28.6 
(Karachentsev et al. 2007) is assumed, with a line-of-sight extinction A$_V = 1.4$. The 
photometric properties of the sources with M$_K \leq -7$ and $J-K < 1.4$ are consistent with 
them belonging to an intermediate age population that has an age $\approx 1$ Gyr 
based on the peak magnitude of AGB stars. The area of the 
LMC $(K, J-K)$ CMD  from Nikolaev \& Weinberg (2000) that contains C stars is also indicated, and a 
number of sources have $J-K$ colors that are consistent with them being C stars.}
\end{figure}

\clearpage

\begin{figure}
\figurenum{9}
\epsscale{0.75}
\plotone{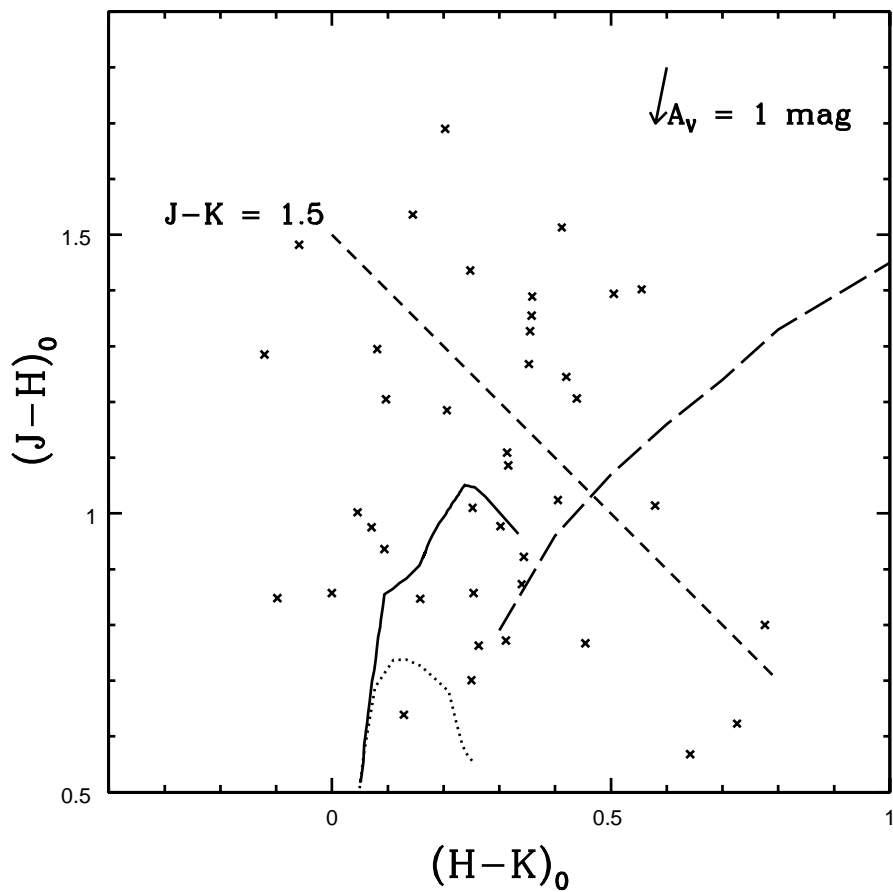}
\caption{The $(H-K, J-H)$ diagram of sources near GSC06726--00126. Only objects with $K < 22$ 
are shown. The loci of solar neighborhood giants and dwarfs from the Z = 0.019 1 Gyr isochrone 
from Girardi et al. (2002) are shown as solid and dotted lines, while the long 
dashed line shows the locus of SMC and LMC LPVs from observations 
tabulated by Wood et al. (1983; 1985). The line indicating $J-K = 1.5$, 
which is the approximate color dividing giants with oxygen-rich and carbon-rich 
atmospheres, is also indicated. The majority of sources with $K < 22$ have $J-H > 0.7$, and hence 
are not foreground main sequence stars.}
\end{figure}

\clearpage

\begin{figure}
\figurenum{10}
\epsscale{0.75}
\plotone{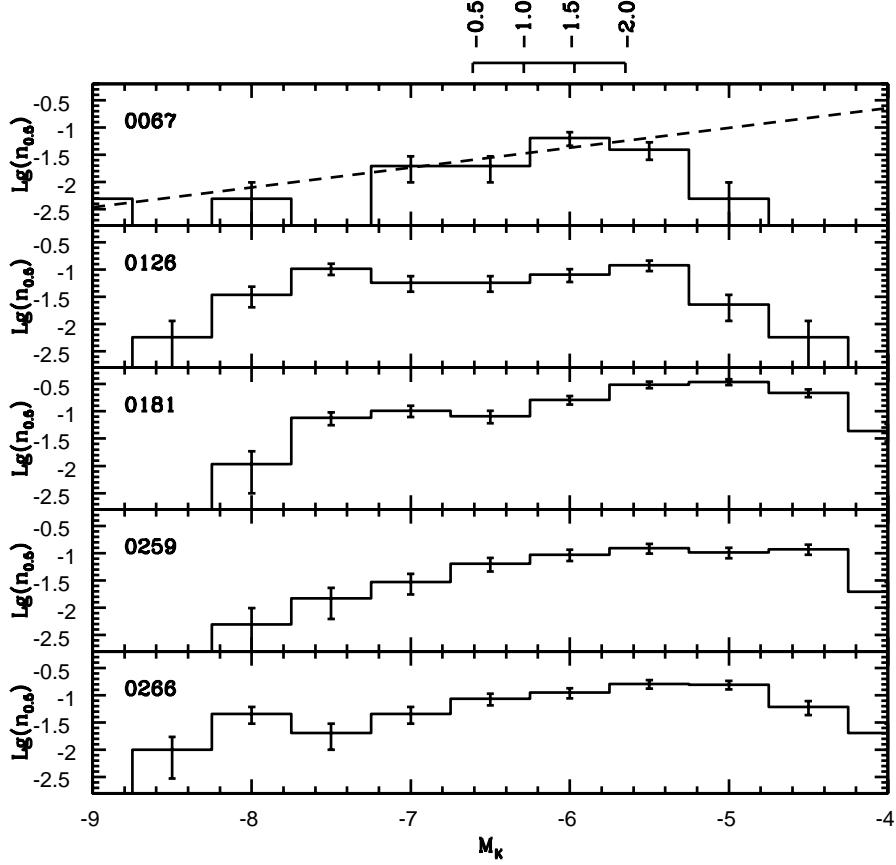}
\caption{The LFs of sources with $(H-K)_0$ between --0.05 and 0.45. n$_{0.5}$ is the number of 
sources arcsec$^{-2}$ per 0.5 magnitude in $K$. The dashed line in the top panel is a least 
squares fit to the GSC06726--0067 LF, and this relation is used to estimate an upper limit to 
background galaxy contamination in the LFs in Figure 11. The RGB-tip brightnesses 
for four metallicities, based on the Galactic globular cluster calibration 
of Ferraro et al. (2000), are indicated at the top of the figure.}
\end{figure}

\clearpage

\begin{figure}
\figurenum{11}
\epsscale{0.75}
\plotone{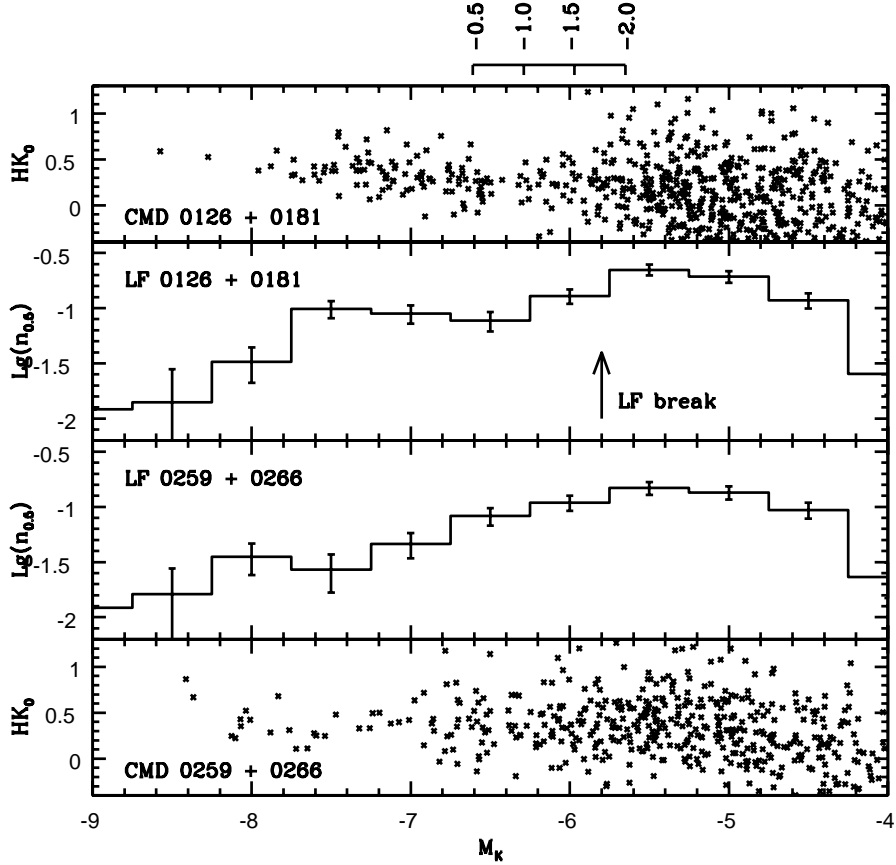}
\caption{The $K$ LFs of sources near GSC06726--0126 $+$ 0181 and GSC06726--0259 $+$ 0266, 
where n$_{0.5}$ is the number of sources arcsec$^{-2}$ per 0.5 magnitude. 
The dashed line in the top panel of Figure 10 was 
subtracted from the summed LFs to correct for contamination from background galaxies. 
The CMDs of GSC06726--0126 $+$ 0181 and GSC06726--0259 $+$ 0266 are also shown to 
permit features in the LFs and CMDs to be compared. The brightness of the RGB-tip for 
systems with various metallicities, based on the calibration of Ferraro et al. (2000), 
is shown at the top of the figure. The break in the mean GSC06726--0126 $+$ 0181 LF 
near M$_K = -5.8$ is accompanied by a broadening of the composite CMD.}
\end{figure}

\clearpage

\begin{figure}
\figurenum{12}
\epsscale{0.75}
\plotone{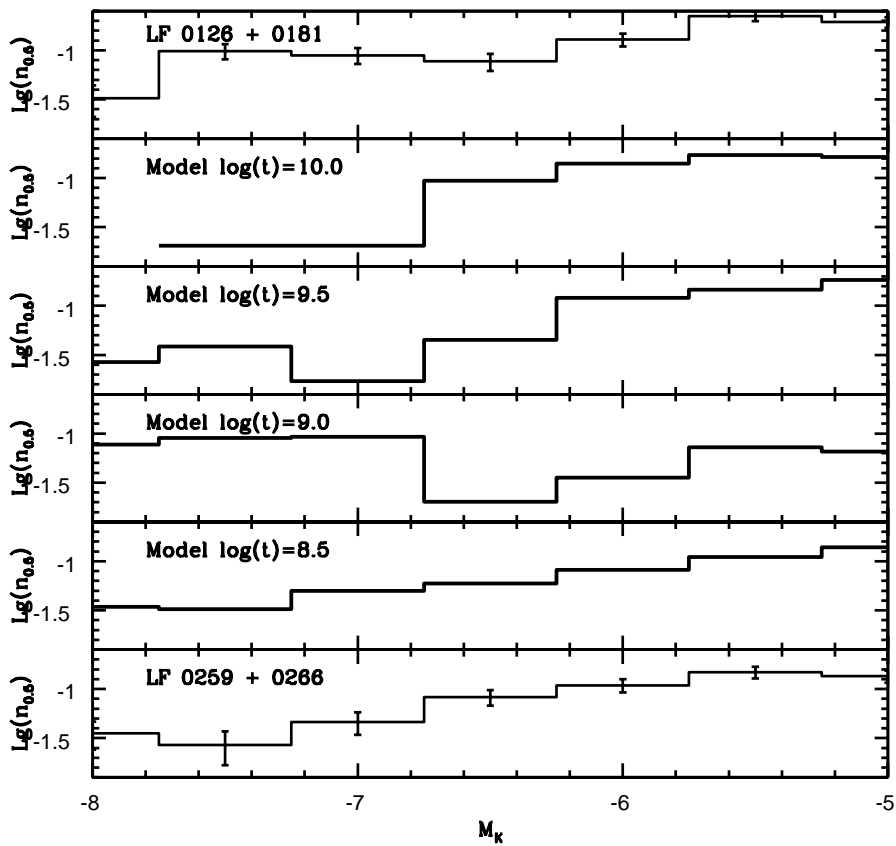}
\caption{The $K$ LFs of sources near GSC06726--0126 $+$ 0181 and GSC06726--0259 $+$ 0266, 
as shown in Figure 11. Model LFs for simple stellar populations with solar 
metallicities constructed from the Girardi et al. (2002) isochrones are also shown. 
The GSC06726--0259$+$0266 LF is well matched by the log(t) = 8.5 LF, whereas the 
GSC06726--0126$+$0181 LF appears to be a mix of the log(t) = 9.0 and older sequences.}
\end{figure}

\clearpage

\begin{figure}
\figurenum{13}
\epsscale{0.75}
\plotone{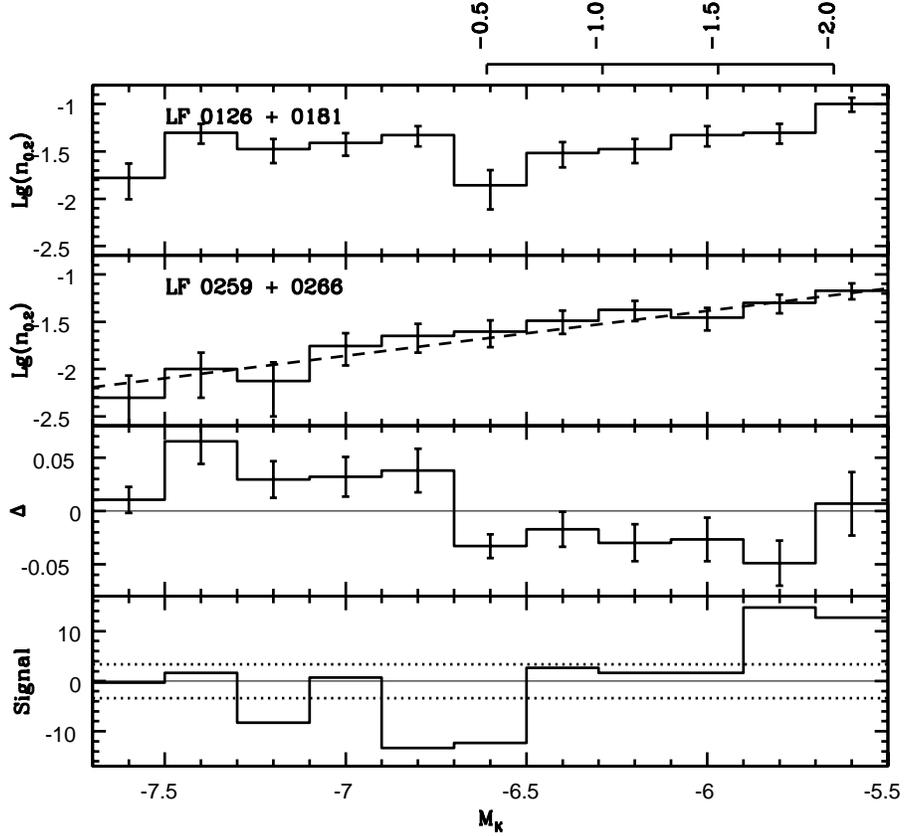}
\caption{The $K$ LFs of sources near GSC06726--0126 $+$ 0181 and GSC06726--0259 $+$ 0266, 
where n$_{0.2}$ is the number of sources arcsec$^{-2}$ per 0.2 magnitude. 
The GSC06726--0259$+$0266 LF follows a power-law, and the 
dashed line in the second panel is a power-law that was fit to the 
LF using the method of least squares. The difference, $\Delta$, between this power-law
and the mean GSC06726--0126$+$0181 LF, after the latter 
was scaled to match the number of stars in the 
GSC06726--0126 and 0181 fields, is shown in the third panel. 
Significant residuals are seen, indicating that the power-law does not represent adequately the 
GSC06726--0126$+$0181 LF. The Sobel-convolved GSC06726--0126$+$0181 LF is shown in 
the bottom panel, and the dotted lines indicate the $1\sigma$ variance in the Sobel-convolved 
GSC06726--0259$+$0266 LF. The break at M$_K = -5.8$ in the 
GSC-06726--0126$+$0266 LF is significant at the $3.6\sigma$ level.
The brightness of the RGB-tip at four metallicities, based on 
the empirical calibration of Ferraro et al. (2000), is shown at the top of the figure. 
This calibration indicates that the break corresponds to [M/H] $= -1.8 \pm 0.1$ 
if the RGB is dominated by old stars. This is lower than expected from a disk population, 
and in \S 3.3 it is argued that the break may not be due to the RGB-tip.} 
\end{figure}

\clearpage

\begin{figure}
\figurenum{14}
\epsscale{0.75}
\plotone{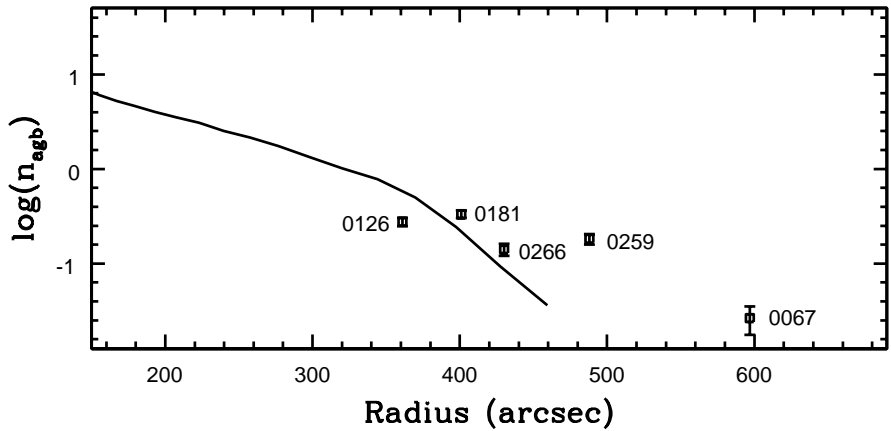}
\caption{The number density of AGB stars as a function of angular separation 
from the center of M83 as measured in the plane of the disk. 
n$_{agb}$ is the number of sources with $K \geq 22.5$ arcsec$^{-2}$, 
and the errorbars show the uncertainties due to counting statistics. 
With the exception of the field around GSC06726--0067, n$_{agb} \approx 0.2$ stars 
arcsec$^{-2}$. The solid line is the number density of AGB stars predicted from the $K-$band 
surface brightness profile of M83 from Jarrett et al. (2003) using the procedure described in 
the text. The densities of AGB stars around GSC06726--0259 and GSC06726--0067 fall well above 
the extrapolation of the number density curve generated from the light profile, 
suggesting that the fractional contribution made by AGB stars to the total light in $K$ 
changes with radius.}
\end{figure}

\end{document}